\documentclass[10pt, conference, compsocconf]{IEEEtran}
\ifCLASSINFOpdf
\else
\fi

\usepackage{algorithm}
\usepackage{algorithmicx}
\usepackage{algpseudocode}
\usepackage{graphics,graphicx}
\usepackage{textcomp}
\usepackage{amsmath}
\usepackage{amsthm}
\usepackage{amssymb}
\usepackage{txfonts}
\usepackage{dsfont}
\usepackage{lipsum}
\usepackage{multicol}
\usepackage{balance}
\usepackage{float}
\usepackage{caption}
\usepackage{setspace}
\usepackage{inputenc}
\usepackage[T1]{fontenc}
\usepackage{cancel}
\usepackage{xcolor,colortbl}
\usepackage{tgcursor}
\usepackage{courier}
\usepackage{subcaption}
\algnewcommand\algorithmicinput{\textbf{Input:}}
\algnewcommand\Input{\item[\algorithmicinput]}

\let\oldReturn\Return
\renewcommand{\Return}{\State\oldReturn}
\algnewcommand\algorithmicoutput{\textbf{Output:}}
\algnewcommand\Output{\item[\algorithmicoutput]}
\newlength\myindent
\algnewcommand\algorithmicforeach{\textbf{for each}}
\algdef{S}[FOR]{ForEach}[1]{\algorithmicforeach\ #1\ \algorithmicdo}

\definecolor{gray}{rgb}{0.86, 0.86, 0.86}

\newtheorem{definition}{Definition}
\newtheorem{theorem}{Theorem}
\newtheorem{lemma}{Lemma}
    \newtheoremstyle{component}{}{}{}{}{\itshape}{.}{.5em}{\thmnote{#3}#1}
    \theoremstyle{component}
    \newtheorem*{component}{}

\begin{document}
%
\title{A Budget Feasible Mechanism for Hiring Doctors in E-Healthcare}

\author{\IEEEauthorblockN{Vikash Kumar Singh}
\IEEEauthorblockA{Dept. of Computer Science and Engineering\\
National Institute of Technology\\
Durgapur 713209, India\\
vikas.1688@gmail.com}\\
\IEEEauthorblockN{Fatos Xhafa}
\IEEEauthorblockA{Dept. of Computer Science\\
Universitat Polit$\grave{e}$cnica de Catalunya\\
Barcelona 08034, Spain\\
fatos@cs.upc.edu}
\and
\IEEEauthorblockN{Sajal Mukhopadhyay}
\IEEEauthorblockA{Dept. of Computer Science and Engineering\\
National Institute of Technology\\
Durgapur 713209, India\\
sajal@cse.nitdgp.ac.in}\\
\IEEEauthorblockN{Aniruddh Sharma}
\IEEEauthorblockA{Dept. of Computer Science and Engineering\\
National Institute of Technology\\
Durgapur 713209, India\\
aniruddhsharma189@gmail.com}}


%


\maketitle

\begin{abstract}
Throughout the past decade, there has been an extensive research on scheduling the hospital resources \emph{such as} the 
operation theatre(s) (OTs)
 and the experts (such as nurses, doctors etc.) inside the hospitals. With the technological growth, mainly advancement in communication media (such as \emph{smart phones}, \emph{video conferencing}, \emph{smart watches} etc.) one may think of taking the expertise by the doctors (distributed around the globe) from outside the in-house hospitals. Earlier this interesting situation of hiring doctors from outside the hospitals has been studied from \emph{monetary} (with patient having \emph{infinite budget}) and \emph{non-monetary} perspectives in \emph{strategic} setting.
In this paper, the more realistic situation is studied in
terms of hiring the doctors from outside the hospital
when a patient is constrained by budget. Our proposed mechanisms follow the two pass mechanism design framework each consisting of \emph{allocation rule} and \emph{payment rule}.
Through simulations, we evaluate the performance
and validate our proposed mechanisms.    
\end{abstract}

%
\IEEEpeerreviewmaketitle

\section{Introduction} 
\begin{figure}[H]
\begin{center}
\includegraphics[scale=0.45]{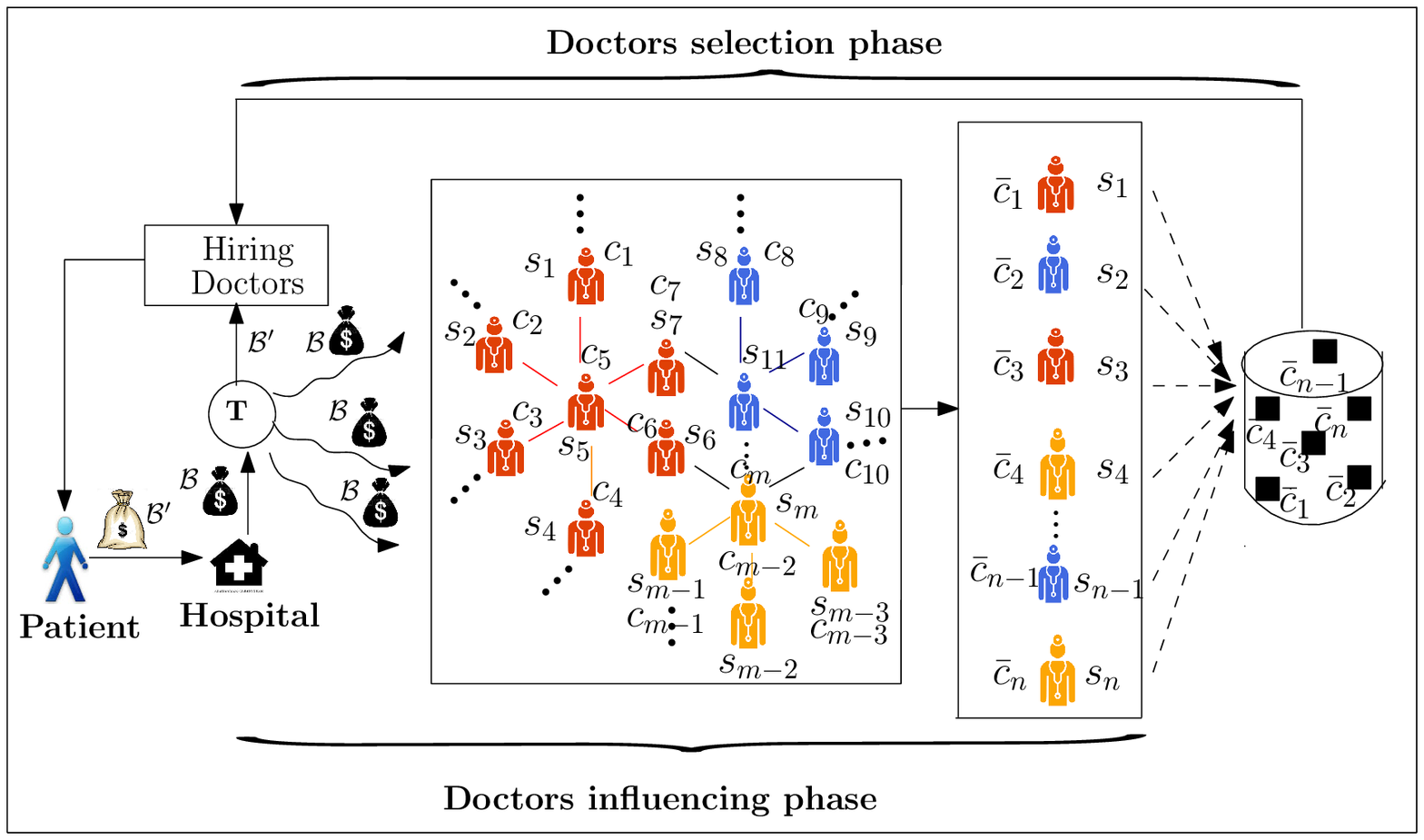}
\caption{System model}
\label{fig:1}
\end{center}
\end{figure}

Substantial literatures are available to schedule resources inside the hospitals in healthcare system \cite{Guerriero_F_2011}\cite{cardoen_b_2010}\cite{Berrada_1996}\cite{Huele_CV_2014}. However, how to hire (to schedule) resources (expert consultants (ECs) etc.) along with their pricing schemes from outside the hospitals are mostly unaddressed \cite{Starren_JB_2014}\cite{Vikash2015EAS}\cite{DBLP:journals/corr/SinghM16}\cite{DBLP:journals/corr/SinghMSR17}. It is observed that, with the
prodigious growth of the communication media (say {\fontfamily{qcr}\selectfont video conferencing},
{\fontfamily{qcr}\selectfont Internet}, {\fontfamily{qcr}\selectfont smart phones} etc.), it may be an usual phenomena to have the consultancies by the experts (especially doctors) from outside the hospital(s). It is to be noted that the doctors can provide consultancies by being present at the consultancy spot (where the patient is admitted) particularly virtually (using {\fontfamily{qcr}\selectfont video conferencing},
{\fontfamily{qcr}\selectfont Internet}, {\fontfamily{qcr}\selectfont smart phones} etc.) etc. making them pervasive. In our future references ECs and doctors will be used interchangeably. In this paper, an attempt is made to hire expert consultants from outside the hospital when a patient is budget limited. The detailing of the hiring concept in this paper is shown in Figure \ref{fig:1}. In our model, the hiring concept is shown as a two fold process. In the first fold, the accumulated hospital's budget can be utilized to detect the leaders in the social graph (representing ECs professional connections) to inform about the hiring concept to the substantial number of doctors.
\indent  
In the second fold, the subset of doctors will be selected from the set consisting of doctors as leaders and the informed doctors for the consultancy process, such that the total payment made to the doctors is within patient's budget.\\
\indent The remainder of the paper is structured as follows. Section II
elucidates the related works. Section III describes our proposed model. The proposed mechanisms are illustrated in section IV. The analysis of the proposed mechanisms are illustrated in section V. In section VI experiments and results are shown. Finally, conclusions are drawn in Section VII.
\section{Related prior work}
In past the handful of works have been done, focusing on scheduling inside the hospitals (or \emph{internal scheduling}) in terms of operation theatres (OTs) scheduling \cite{Chan:2016:PCP:2936924.2936967}\cite{Guerriero_F_2011}\cite{cardoen_b_2010}\cite{Cardoen_B_1_2010} and internal staffs (such as \emph{nurses} \cite{Berrada_1996}\cite{KO_Y_W_2017}, \emph{physicians} \cite{Carter_2001}\cite{Huele_CV_2014} etc.) scheduling. In \cite{Dexter_F_2004}\cite{Cardoen_B_1_2010} the works have been done for allocating OTs on time to increase OTs efficiency. In our future references, hospital(s), medical unit(s), organization(s) will be used interchangeably. As with the enhancement in the technologies, mainly communication media (say video conferencing,
Internet, smart phones etc.), it may be an usual phenomena to think of the scheduling of medical staffs (\emph{mainly doctors}) outside the in-house hospital \cite{Starren_JB_2014}\cite{Vikash2015EAS}\cite{7920909}. In \cite{Starren_JB_2014} a doctor is providing the expertise through video conferencing to a patient admitted to other hospital with prior contact. In \cite{TETC_2014_2386133} the context of the patient (such as \emph{age}, \emph{sex}, \emph{medical report} etc.) is utilized to take the expertise of the doctors from outside the admitted hospitals in non-strategic setting. In \cite{Vikash2015EAS} the \emph{strategic} case is considered and is solved using mechanism design with money and in \cite{DBLP:journals/corr/SinghM16}\cite{DBLP:journals/corr/SinghMSR17} mechanism design without money is utilized. Despite some progress in the scenario of hiring ECs from outside the hospital(s), the patients with budget constraints case has been largely overlooked. In this paper, the problem of hiring doctors from outside the hospital is studied in this setting.

\section{System model}
\subsection{Notation and Preliminaries}
In this section, we formalize the doctors hiring problem where the multiple doctors are hired from outside of the hospital, for a patient having budget $\boldsymbol{\mathcal{B}'}$. The patient's budget $\boldsymbol{\mathcal{B}'}$ is a \emph{public} information. The hospital to which a patient is admitted is having an accumulated, publicly known budget $\boldsymbol{\mathcal{B}}$, which will be utilized to inform about the hiring concept to the substantial number of ECs. The set of ECs is given as $\mathcal{S} = \{s_1, s_2, \ldots, s_m\}$; where each EC $s_i \in \mathcal{S}$ is assumed to be professionally connected with some $\chi_{i} \subseteq \mathcal{S}$$\setminus$$\{s_i\}$. The professional connections are given by a social graph $\mathcal{\mathcal{G}(\mathcal{V}, \mathcal{E})}$, where $\mathcal{V}$ is the set of nodes representing ECs and $\mathcal{E}$ is the set of edges representing their professional connections in the social graph. Each $s_i$ is associated with a hospital $\hbar_i \in \boldsymbol{\mathcal{H}}$.
Our model consists of two fold process. In the {\fontfamily{qcr}\selectfont first fold}, there is a social graph that is represented as $\mathcal{\mathcal{G}(\mathcal{V}, \mathcal{E})}$ and publicly known \emph{expert consultant activation function} given as $\boldsymbol{\mathcal{I}}: 2^{\mathcal{S}} \rightarrow \boldsymbol{\mathcal{R}_{\geq 0}}$. Given the subset $\varGamma \subseteq \mathcal{S}$ the value $\mathcal{I}(\varGamma)$ represents the expected number of doctors that are made aware about the hiring concept $i.e.$ $I(\varGamma) = |\underset{i \in \varGamma}{\bigcup} \chi_i|$. Each node in the graph represents a doctor $s_i$ that has a private cost (\emph{aka} bid) $c_i$ of being an initial adapter or the cost for spreading awareness about the hiring concept to other doctors. The cost vector of all the $m$ doctors is given as: $\boldsymbol{\mathcal{C}} = \{c_1, c_2, \ldots, c_m\}$. It is to be noted that, the ECs are {\fontfamily{qcr}\selectfont rational} and {\fontfamily{qcr}\selectfont strategic} in nature. It means that, the ECs can gain by misreporting their private cost. As the ECs are {\fontfamily{qcr}\selectfont strategic}; so each $s_i \in \mathcal{S}$ may report their cost for being an initial adapter as $c'_i$ instead of $c_i$ in order to gain; where $c'_i \neq c_i$. The payment vector for the set $\varGamma$ is given as $\boldsymbol{\mathcal{P}}_{\varGamma}$ = $\{\boldsymbol{\mathcal{P}}_{\varGamma_1}, \boldsymbol{\mathcal{P}}_{\varGamma_2}, \ldots, \boldsymbol{\mathcal{P}}_{\varGamma_k}\}$; where $\boldsymbol{\mathcal{P}}_{\varGamma_i}$ is the payment of $s_i \in \varGamma$.  
 The objective of the first fold is to maximize the \emph{expert consultant activation function} while the total payment is at most hospital's budget $\boldsymbol{\mathcal{B}}$.    
In the {\fontfamily{qcr}\selectfont second fold}, we have a set of doctors consisting of doctors acted as leaders in the \emph{first fold} and the aware doctors given as $\hat{\mathcal{S}} = \{s_1, s_2, \ldots, s_{i-1}, s_{i}, \ldots, s_{n}\}$ such that $n \leq m$. The quality vector of all the $m$ ECs is given as $\boldsymbol{\mathcal{Q}} = \{\mathcal{Q}_1, \mathcal{Q}_2, \ldots, \mathcal{Q}_m\}$, where $\mathcal{Q}_i \in \boldsymbol{\mathcal{Q}}$ is the quality of $i^{th}$ doctor. In general, the quality $\mathcal{Q}_{i}$ of a doctor $s_i$ can be estimated using various parameters calculated later in the section. The publicly known \emph{quality} function is given as $\boldsymbol{\mathcal{D}}: 2^{\hat{\mathcal{S}}} \rightarrow \boldsymbol{\mathcal{R}_{\geq 0}}$. Given a subset $\varUpsilon \subseteq \hat{\mathcal{S}}$, the value $\boldsymbol{\mathcal{D}}(\varUpsilon)$ denotes the sum of the qualities of all the doctors in $\varUpsilon$  $i.e.$ $\boldsymbol{\mathcal{D}}(\varUpsilon) = \sum_{i \in \varUpsilon} \mathcal{Q}_{i}$. For this fold, each doctor $s_i \in \hat{\mathcal{S}}$ will bid afresh their cost (\emph{private}) for providing consultancy to the patient and is given as $\bar{c}_i$. The cost vector of all the \emph{n} doctors is given as: $\boldsymbol{\mathcal{\bar{C}}} = \{\bar{c}_1, \bar{c}_2, \ldots, \bar{c}_n\}$. The {\fontfamily{qcr}\selectfont strategic} behaviour of the doctors is continued in this fold also; so each $s_i \in \hat{\mathcal{S}}$ may report their cost of consultancy as $\bar{c}'_i$ instead of $\bar{c}_i$ in order to gain; where $\bar{c}'_i \neq \bar{c}_i$.  
Our objective is to determine the subset $\varUpsilon \in \{\xi|\sum_{i\in \xi} \bar{c}_{i} \leq \boldsymbol{\mathcal{B}}' \}$ for which $\boldsymbol{\mathcal{D}}(\varUpsilon)$ is maximized and the total payment should not exceed the patient's budget $\boldsymbol{\mathcal{B}}'$.  The payment vector of the set $\varUpsilon$ is given as $\hat{\boldsymbol{\mathcal{P}}}$ = $\{\hat{\boldsymbol{\mathcal{P}}}_1$, $\hat{\boldsymbol{\mathcal{P}}}_2$, \ldots, $\hat{\boldsymbol{\mathcal{P}}}_r\}$.

\subsection{Quality determination}
The parameters that determine the quality of each doctor $s_i$ are: (1) {\fontfamily{qcr}\selectfont qualification} of $s_i$ given as $q_i$ (2) {\fontfamily{qcr}\selectfont success rate} of $s_i$ given as $sr_i$ (3) {\fontfamily{qcr}\selectfont experience} of $s_i$ given as $e_i$ (4) {\fontfamily{qcr}\selectfont hospital} to which $s_i$ belong given as $\hbar_i$. So, the quality of doctor $s_i$ is given as: $\mathcal{Q}_i=\left(w_1\cdot q_i + w_2 \cdot sr_{i} + w_3 \cdot e_i + w_4 \cdot \hbar_i\right)$; where, $w_i \in [0,1]$ such that $\sum_{i} w_i =1$. The weighted sum of the some of the parameters considered in our case will result in the quality of the doctors. 
\subsection{Budget distribution and utilization} 
In our scenario, each fold is utilizing the budget from two independent sources. Firstly, talking about the hospital's budget it can be thought of as 1) the accumulated fund from the previously admitted patients say adding 5-6\% of the total fees of each patients to the hospital fund. 2) Donation to the hospital by high profile persons or communities. Next, the source of the budget utilized in the second phase is the patient itself.
 \begin{definition}[\emph{Marginal Contribution} \cite{Singer:2012:WFI:2124295.2124381}]
The marginal contribution of an EC $s_i \in \mathcal{S}$ is the number of ECs informed about the hiring concept by the EC $s_i$ given the set of $i-1$ ECs $i.e.$ $\varGamma_{i-1}$ already selected as the leaders. 
Mathematically, the marginal contribution of $i^{th}$ EC given $\varGamma_{i-1}$ is defined as: $\boldsymbol{\mathcal{M}}_{\boldsymbol{\mathcal{C}}_i}(\varGamma_{i-1}) = \boldsymbol{\mathcal{I}}(\varGamma_{i-1} \cup \{s_i\}) - \boldsymbol{\mathcal{I}}(\varGamma_{i-1})$
 
\end{definition} 

 \begin{definition}[\emph{Quality Contribution} \cite{Singer:2014:BFM:2692359.2692366}]
The quality contribution of an EC $s_i \in \mathcal{S}$ given a subset $\varUpsilon_{i-1}$ of ECs already been selected is given as: $\boldsymbol{\mathcal{D}}_{i}(\varUpsilon_{i-1}) = \boldsymbol{\mathcal{D}}(\varUpsilon_{i-1} \cup \{s_i\})- \boldsymbol{\mathcal{D}}(\varUpsilon_{i-1})$ where $\boldsymbol{\mathcal{D}}(\varUpsilon_{i})$ denotes the sum of the qualities of all the doctors in $\varUpsilon_{i}$  $i.e.$ $\boldsymbol{\mathcal{D}}(\varUpsilon_{i}) = \sum_{i} \mathcal{Q}_{i}$ given $\varUpsilon_{i} = \{1, \ldots, i\}$ and $\boldsymbol{\mathcal{D}}(\varUpsilon_0) = 0$ as $\varUpsilon_0 = \phi$.
\end{definition} 
\section{Proposed mechanisms}
In this section, we present proposed mechanisms: \emph{\textbf{No}n-\textbf{t}ruthful \textbf{b}udget \textbf{c}onstraint (NoTBC) mechanism} motivated by \cite{Khuller:1999:BMC:332530.332554} and \emph{\textbf{T}ruthful \textbf{b}udget \textbf{c}onstraint (TBC) mechanism} motivated by \cite{Singer:2012:WFI:2124295.2124381}\cite{Singer:2014:BFM:2692359.2692366}. 
\subsection{NoTBC mechanism}
It is a two pass mechanism consisting of \emph{{\textbf{No}n-\textbf{t}ruthful \textbf{b}udget \textbf{c}onstraint \textbf{l}eader \textbf{i}dentification (NoTBC-LI) mechanism}} and \emph{{\textbf{No}n-\textbf{t}ruthful \textbf{b}udget \textbf{c}onstraint \textbf{d}octor \textbf{s}election (NoTBC-DS) mechanism}}. 
\subsubsection{NoTBC-LI mechanism}
In each iteration of \emph{while} loop, a doctor with maximum marginal contribution per cost among the available doctors is considered and is selected only if its cost for being an initial adapter is less than the hospital's available budget. The payment of each doctors as a leader is their revealed cost. 
\begin{algorithm}[H]
\caption{NoTBC-LI mechanism ($\mathcal{G}$, $\mathcal{S}$, $\boldsymbol{\mathcal{B}}$, $\boldsymbol{\mathcal{C}}$)}
\begin{algorithmic}[1]
      \Output $\hat{\mathcal{S}}$ $\leftarrow$ $\phi$, $\boldsymbol{\mathcal{P}}_{\varGamma} \leftarrow \phi$
	\State $\bar{\mathcal{S}}$ $\leftarrow$ $\phi$ \Comment{Set containing all the informed doctors.}
	      \While {$\mathcal{S} \neq \phi$}
	      \State $s_i$ $\leftarrow$ $argmax_{j \in \mathcal{S}}$ $\left[\frac{\boldsymbol{\mathcal{M}}_{\boldsymbol{\mathcal{C}}_{j}}(\varGamma_{j-1})}{c_j}\right]$
	        \If{$c_i \leq \boldsymbol{\mathcal{B}}$ }
	              \State $\varGamma$ $\leftarrow$ $\varGamma$ $\cup$ $\{s_i\}$; $\bar{\mathcal{S}}$ $\leftarrow$ $\bar{\mathcal{S}}$ $\cup$ $\{\chi_{i}\}$; $\boldsymbol{\mathcal{B}} \leftarrow \boldsymbol{\mathcal{B}} - c_i$ 
	        \EndIf 
	        \State $\mathcal{S} \leftarrow \mathcal{S} \setminus \{s_i\}$     
	      \EndWhile
	      \State $\hat{\mathcal{S}} = \varGamma \cup \bar{\mathcal{S}}$
	 \For{$each$ $s_i$ $\in$ $\varGamma$}
	     \State $\boldsymbol{\mathcal{P}}_{\varGamma_i} \leftarrow c_i$; $\boldsymbol{\mathcal{P}}_{\varGamma} \leftarrow \boldsymbol{\mathcal{P}}_{\varGamma} \cup \{\boldsymbol{\mathcal{P}}_{\varGamma_i}\}$ 
	 \EndFor     
	      \Return $\hat{\mathcal{S}}$, $\boldsymbol{\mathcal{P}}_{\varGamma}$   
 
\end{algorithmic}
\end{algorithm} 
\subsubsection{NoTBC-DS mechanism}
In each iteration of \emph{while} loop, a doctor with maximum quality contribution per cost among the selected doctors by {\fontfamily{qcr}\selectfont NoTBC-LI mechanism} is considered and is hired only if its cost for the consultancy is less than the patient's available budget. The payment of each hired doctors is their revealed cost of consultancy. 
\begin{lemma}
The NoTBC mechanism is computationally efficient. 
\end{lemma}
\begin{proof}
In {\fontfamily{qcr}\selectfont NoTBC-LI}, for \emph{m} iteration of \emph{while} loop  we have $O(m^2)$. Thus, the running time of {\fontfamily{qcr}\selectfont NoTBC-LI} is $O(m^2)$. In {\fontfamily{qcr}\selectfont NoTBC-DS}, for $m$ iteration (in worst case) of \emph{while} loop we have $O(m^2)$. Thus, the running time of {\fontfamily{qcr}\selectfont NoTBC-DS} is $O(m^2)$. In both the cases, the payment determination will be linear in \emph{m}. Thus, the computational complexity of {\fontfamily{qcr}\selectfont NoTBC} is given as $O(m^2)$.
\end{proof}
\begin{algorithm}[H]
\caption{NoTBC-DS mechanism ($\hat{\mathcal{S}}$, $\boldsymbol{\mathcal{B}}'$, $\boldsymbol{\mathcal{\bar{C}}}$)}
\begin{algorithmic}[1]
      \Output $\varUpsilon$ $\leftarrow$ $\phi$, $\hat{\boldsymbol{\mathcal{P}}} \leftarrow \phi$
	      \While {$\hat{\mathcal{S}} \neq \phi$}
	      \State $s_i$ $\leftarrow$ $argmax_{j \in \hat{\mathcal{S}}}$ $\frac{\boldsymbol{\mathcal{D}}_{j}(\varUpsilon_{j-1})}{\bar{c}_j}$
	        \If{$\bar{c}_i \leq \boldsymbol{\mathcal{B}'}$ }
	              \State $\varUpsilon$ $\leftarrow$ $\varUpsilon$ $\cup$ $\{s_{i}\}$; $\boldsymbol{\mathcal{B}'} \leftarrow \boldsymbol{\mathcal{B}'} - \bar{c}_i$
	        \EndIf 
	        \State $\hat{\mathcal{S}} \leftarrow \hat{\mathcal{S}} \setminus \{s_i\}$     
	      \EndWhile
	\For{$each$ $s_i$ $\in$ $\varUpsilon$}    
	       \State $\hat{\boldsymbol{\mathcal{P}}}_i \leftarrow \bar{c}_i$; $\hat{\boldsymbol{\mathcal{P}}} \leftarrow \hat{\boldsymbol{\mathcal{P}}} \cup \{\hat{\boldsymbol{\mathcal{P}}}_i\}$
	\EndFor
	      \Return $\varUpsilon$, $\hat{\boldsymbol{\mathcal{P}}}$      
 
\end{algorithmic}
\end{algorithm}

\begin{lemma}
The NoTBC mechanism is individually rational. 
\end{lemma}
\begin{proof}
 From line 11 of Algorithm 1, we can see $\boldsymbol{\mathcal{P}}_{\varGamma_i}$ = $c_i$ for each $s_i \in \varGamma$. Line 9 in Algorithm 2 shows that $\hat{\boldsymbol{\mathcal{P}}}_i$ = $\bar{c}_i$. Therefore, we have payment for any winner is its cost. Hence, NoTBC mechanism is individually rational.   
\end{proof}

\begin{lemma}
The NoTBC mechanism is budget feasible. 
\end{lemma}
\begin{proof}
 As it is clear that a doctor is included in the winning set only when the given condition in line 4 of Algorithm 1 and line 3 of Algorithm 2 is satisfied. As the payment in case of NoTBC is equal to the cost; the total payment will be at most the budget. Hence, NoTBC mechanism is budget feasible. 
\end{proof}
\subsection{TBC mechanism}
It is a two pass mechanism consisting of \emph{{\textbf{T}ruthful \textbf{b}udget \textbf{c}onstraint \textbf{l}eader \textbf{i}dentification (TBC-LI)}} and \emph{{\textbf{T}ruthful \textbf{b}udget \textbf{c}onstraint \textbf{d}octor \textbf{s}election (TBC-DS)}} mechanisms. 
\subsubsection{TBC-LI mechanism}
\begin{figure}[H]
        \begin{subfigure}[b]{0.5\linewidth}
        \centering
                \includegraphics[width=2.7cm,height=2.7cm]{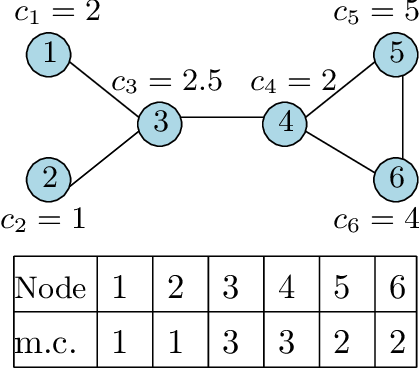}
                \caption{Initial configuration}
                \label{fig:gull}
        \end{subfigure}%
        ~~~
        \begin{subfigure}[b]{0.5\linewidth}
        \centering
                \includegraphics[width=2.7cm,height=2.7cm]{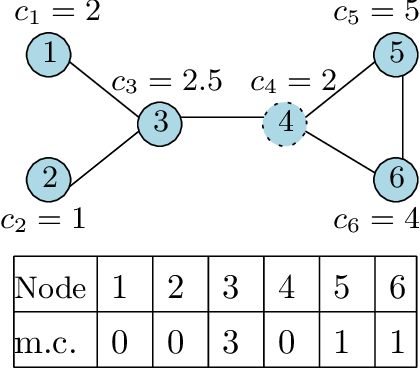}
                \caption{Intermediate configuration}
                \label{fig:gull2}
        \end{subfigure}
        \caption{Detailed functioning of Algorithm 3}\label{fig:animals}
\end{figure}

\begin{figure}[H]
\begin{subfigure}{.5\linewidth}
\centering
\includegraphics[width=2.7cm,height=2.7cm]{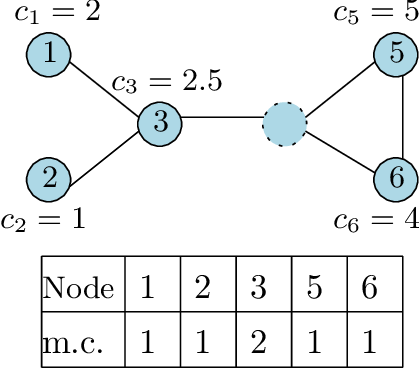}
\caption{Initial configuration}
\label{fig:sub1}
\end{subfigure}%
\begin{subfigure}{.5\linewidth}
\centering
\includegraphics[width=2.7cm,height=2.7cm]{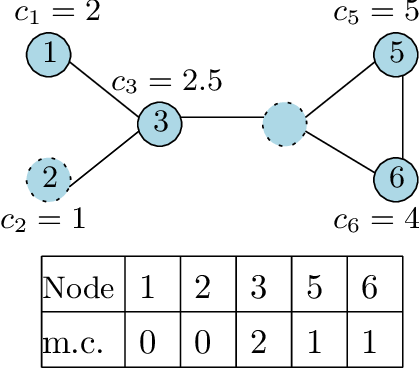}
\caption{Intermediate configuration}
\label{fig:sub2}
\end{subfigure}
~~
\begin{subfigure}{.5\linewidth}
\centering
\includegraphics[width=2.7cm,height=2.7cm]{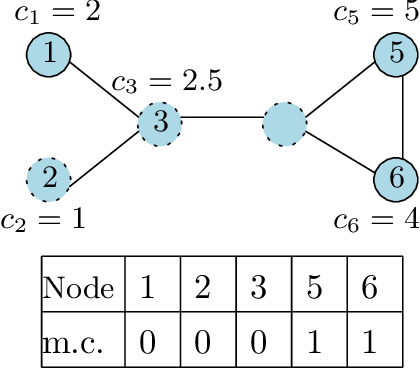}
\caption{Intermediate configuration}
\label{fig:sub3}
\end{subfigure}%
\begin{subfigure}{.5\linewidth}
\centering
\includegraphics[width=3cm,height=2.7cm]{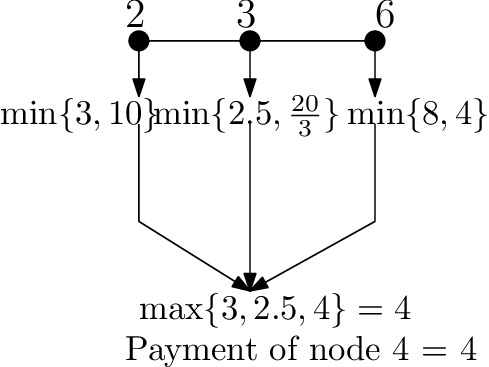}
\caption{Payment determination}
\label{fig:sub4}
\end{subfigure}
\caption{Payment calculation of node 4}
\label{fig:test}
\end{figure}

For first fold of hiring problem, we propose a TBC-LI mechanism motivated by \cite{Singer:2012:WFI:2124295.2124381}\cite{Singer:2014:BFM:2692359.2692366}. 
\paragraph{\textbf{Allocation rule}}
In this, a doctor with maximum marginal contribution per cost among the available doctors
\begin{algorithm}[H]
\caption{TBC-LI allocation mechanism ($\mathcal{G}$, $\mathcal{S}$, $\boldsymbol{\mathcal{B}}$, $\boldsymbol{\mathcal{C}}$)}
\begin{algorithmic}[1]
      \Output  $\varGamma \leftarrow \phi$, $\hat{\mathcal{S}}$ $\leftarrow$ $\phi$
  
	\State $\bar{\mathcal{S}} \leftarrow \phi$; $s_i$ $\leftarrow$ $argmax_{j \in \mathcal{S}}$ $\left[\frac{\boldsymbol{\mathcal{M}}_{\boldsymbol{\mathcal{C}}_{j}}(\varGamma_{j-1})}{c_j}\right]$
	
	      \While {$c_i$ $\leq$ $\frac{\boldsymbol{\mathcal{B}}}{2}$ $ \left( \frac{\boldsymbol{\mathcal{M}}_{\boldsymbol{\mathcal{C}}_{i}}(\varGamma_{i-1})}{\boldsymbol{\mathcal{M}}_{\boldsymbol{\mathcal{C}}_{i}}(\varGamma_{i-1}) + \boldsymbol{\mathcal{I}}(\varGamma_{i-1})}\right)$}
	       \State $\varGamma$ $\leftarrow$ $\varGamma$ $\cup$ $\{s_i\}$; $\bar{\mathcal{S}}$ $\leftarrow$ $\bar{\mathcal{S}}$ $\cup$ $\{\chi_{i}\}$ 
	      \State $s_{i}$ $\leftarrow$ $argmax_{j \in \mathcal{S}\setminus \varGamma} \left[\frac{\boldsymbol{\mathcal{M}}_{\boldsymbol{\mathcal{C}}_{j}}(\varGamma_{j-1})}{c_j}\right]$
	      \EndWhile
	       \State $\hat{\mathcal{S}} = \varGamma \cup \bar{\mathcal{S}}$
	      \Return $\varGamma$ and $\hat{\mathcal{S}}$   
 
\end{algorithmic}
\end{algorithm}
 is considered. But the doctor is selected as the leader only when the ratio between their cost as the initial adapter and budget is less than or equal to half of the ratio between their
marginal contribution and the value of the selected subset.\\  
\indent \textbf{\underline{Example 1(a):}}
Figure \ref{fig:gull} show the initial configuration of the social graph along with cost distribution, and marginal contribution (m.c.). The quality vector of the nodes is given as: $\boldsymbol{\mathcal{Q}}=\{5,~ 1,~ 3,~ 5,~ 4,~ 5\}$. Higher the value, higher will be the quality. For understanding purpose we are taking the quality of the doctors as an integer value but in general it may not be the case. It is to be noted that the unit of cost and budget is taken as \$. We have considered hospital's budget to be 10.   
Using line 1 of the {\fontfamily{qcr}\selectfont Algorithm 3} the node 4 is considered. 
\noindent The condition $2 \leq 5 \cdotp (\frac{3}{3+0})$ for node 4 is satisfied. So, $\varGamma = \{4\}$ and $\bar{\mathcal{S}} = \{3,~5,~6\}$. Next, node 3 will be considered and $2.5 \leq 5 \cdotp (\frac{3}{3+3})$ for node 3 is satisfied. So, $\varGamma = \{4,~3\}$ and $\bar{\mathcal{S}} = \{3,~5,~6,~1,~2,~4\}$.  So, we have $\hat{\mathcal{S}} = \{3,~5,~6,~1,~2,~4\}$. 
  \paragraph{\textbf{Payment rule}}
The payment rule is motivated by \cite{Singer:2014:BFM:2692359.2692366}.
  \begin{algorithm}[H]
\caption{TBC-LI pricing mechanism ($\varGamma$, $\boldsymbol{\mathcal{B}}$, $\boldsymbol{\mathcal{C}}$)}
\begin{algorithmic}[1]
      \Output $\boldsymbol{\mathcal{P}}_{\varGamma}$ $\leftarrow$ $\phi$.
	\State $\varGamma' \leftarrow \phi$, $\mathcal{S}' \leftarrow \phi$
	\For{$each$ $s_i$ $\in$ $\varGamma$}
	    \State $\mathcal{S}'$ $\leftarrow$ $\mathcal{S} \setminus \{s_i\}$  
	    \State $s_j$ $\leftarrow$ $argmax_{k \in \mathcal{S}'}$ $\left[\frac{\boldsymbol{\mathcal{M}}_{\boldsymbol{\mathcal{C}}_{k}}(\varGamma_{k-1}')}{c_k}\right]$
	    \While {$c_j$ $\leq$ $\boldsymbol{\mathcal{B}} \left( \frac{\boldsymbol{\mathcal{M}}_{\boldsymbol{\mathcal{C}}_{j}}(\varGamma_{j-1}')}{\boldsymbol{\mathcal{M}}_{\boldsymbol{\mathcal{C}}_{j}}(\varGamma_{j-1}') + \boldsymbol{\mathcal{I}}(\varGamma_{j-1}')}\right)$}
	    \State $\varGamma'$ $\leftarrow$ $\varGamma'$ $\cup$ $\{s_j\}$ \Comment{$\varGamma'$ is the set of leaders when $s_i$ is not in the market.}
 	    \State $\mathcal{S}'$ $\leftarrow$ $\mathcal{S}' \setminus \{s_j\}$
	    \State $s_j$ $\leftarrow$ $argmax_{k \in \mathcal{S}'}$ $\left[\frac{\boldsymbol{\mathcal{M}}_{\boldsymbol{\mathcal{C}}_{k}}(\varGamma_{k-1}')}{c_k}\right]$
	    \EndWhile
           \State $\varGamma' \leftarrow \varGamma' \cup \{s_j\} $
	    \For{$each$ $s_j$ $\in$ $\varGamma'$}
	      \State Calculate $\boldsymbol{\mathcal{C}}_{i}^{j}$ = $\frac{\boldsymbol{\mathcal{M}}_{\boldsymbol{\mathcal{C}}_{i}}^{j}(\varGamma_{j-1}') \cdot c_{j}}{\boldsymbol{\mathcal{M}}_{\boldsymbol{\mathcal{C}}_{j}}(\varGamma_{j-1}')}$ and $\boldsymbol{\Pi}_{i}^{j} = \frac{\boldsymbol{\mathcal{B}} \cdot \boldsymbol{\mathcal{M}}_{\boldsymbol{\boldsymbol{\mathcal{C}}}_i}^{j}(\varGamma_{j-1}')}{\boldsymbol{\mathcal{I}}(\varGamma_{j-1}' \cup \{s_i\})}$
	      
	      \EndFor
	      \State $\boldsymbol{\mathcal{P}}_{\varGamma_i} \leftarrow max_{j \in [1 \ldotp \ldotp \ell + 1]}\{min\{\boldsymbol{\mathcal{C}}_{i}^{j}, \boldsymbol{\Pi}_{i}^{j}\}\}$; $\boldsymbol{\mathcal{P}}_{\varGamma}$ $\leftarrow$ $\boldsymbol{\mathcal{P}}_{\varGamma} \cup \{\boldsymbol{\mathcal{P}}_{\varGamma_i}\}$    
	\EndFor
	      \Return $\boldsymbol{\mathcal{P}}_{\varGamma}$     
\end{algorithmic}
\end{algorithm}
 In this, for each doctor $s_i \in \varGamma$ consider running line $3-9$.
 Next, determine the largest index $\ell$ in the sorting of $|\varGamma'|$ doctors (determined without $s_i$) such that the ratio between their cost as the initial adapter and budget is less
than or equal to the ratio between their marginal contribution and
the value of the selected subset. Now for each point $j \in [1 \ldotp \ldotp \ell+1 ]$ find the maximal cost $\boldsymbol{\mathcal{C}}_{i}^{j} = \boldsymbol{\mathcal{M}}_{\boldsymbol{\mathcal{C}}_{i}}^{j}(\varGamma_{j-1}') \cdot \bigg(\frac{c_{j}}{\boldsymbol{\mathcal{M}}_{\boldsymbol{\mathcal{C}}_{j}}(\varGamma_{j-1}')}\bigg)$ that doctor $s_i$ can declare in order to be allocated instead of the doctor in the $j^{th}$ place in the sorting; where $\boldsymbol{\mathcal{M}}_{\boldsymbol{\mathcal{C}}_{i}}^{j}(\varGamma_{j-1}')$ is the marginal contribution of doctor $s_i$ when considered on $j^{th}$ place is given as:  $\boldsymbol{\mathcal{M}}_{\boldsymbol{\mathcal{C}}_{i}}^{j}(\varGamma_{j-1}') = \boldsymbol{\mathcal{I}}(\varGamma_{j-1}' \cup \{s_i\}) - \boldsymbol{\mathcal{I}}(\varGamma_{j-1}')$. Now, if this cost does not exceed the threshold payment $\boldsymbol{\Pi}_{i}^{j} = \boldsymbol{\mathcal{B}} \cdot \frac{\boldsymbol{\mathcal{M}}_{\boldsymbol{\boldsymbol{\mathcal{C}}}_i}^{j}(\varGamma_{j-1}')}{\boldsymbol{\mathcal{I}}(\varGamma_{j-1}' \cup \{s_i\})}$ then the mechanism will declare $s_i$ as the leader. Considering the maximum of the values at $j \in [1 \ldotp \ldotp \ell+1]$ results in the payment of $s_i$.\\ 
\indent \textbf{\underline{Example 1(b):}}
Figure \ref{fig:test} shows the payment calculation of node 4. So, placing node 4 outside the market and utilizing line $3-9$ of Algorithm 4 on the configurations shown in Figure \ref{fig:sub1}, Figure \ref{fig:sub2}, and Figure \ref{fig:sub3} we find the critical point as $\ell=2$ (index of node 3). 
Following Figure \ref{fig:sub4} at point 1 (index of node 2) the value $\boldsymbol{\mathcal{C}}_{4}^{1} = 3 \cdot (\frac{1}{1})=3$, and $\boldsymbol{\Pi}_{4}^{1} = 10 \cdot (\frac{3}{3})=10$. So, $\min\{3,10\}=3$. Similarly, at point 2 (index of node 3) the value $\boldsymbol{\mathcal{C}}_{4}^{2} = 2 \cdot (\frac{2.5}{2})=2.5$, and $\boldsymbol{\Pi}_{4}^{2} = 10 \cdot (\frac{2}{3})=6.66$. So, $\min\{2.5,6.66\}=2.5$.
Considering point 3 (index of the first loser node $i.e$ node 6) we get $\boldsymbol{\mathcal{C}}_{4}^{3} = 2 \cdot (\frac{4}{1})=8$, and $\boldsymbol{\Pi}_{4}^{3} = 10 \cdot (\frac{2}{5})=4$. So, $\min\{8,~4\}=4$. The payment of node 4 is $\max\{3,~2.5,~4\}=4$. 
\subsubsection{TBC-DS mechanism}
For the second fold of the doctors hiring problem, we propose a {\fontfamily{qcr}\selectfont TBC-DS mechanism} motivated by \cite{Singer:2012:WFI:2124295.2124381}\cite{Singer:2014:BFM:2692359.2692366}. 
\paragraph{\textbf{Allocation rule}}
In this, firstly the available doctors are sorted in decreasing order based on quality contribution by cost. Now, the doctors are greedily selected but will be hired only when the ratio of the selected doctor's cost of consultation and the patient's budget is less than or equal to the ratio between the quality contribution by the selected doctor and the value of the quality of the selected  subset.
 \begin{algorithm}[H]
\caption{TBC-DS allocation mechanism ($\hat{\mathcal{S}}$, $\boldsymbol{\mathcal{B}}'$, $\bar{\boldsymbol{\mathcal{C}}}$)}
\begin{algorithmic}[1]
      \Output $\varUpsilon$ $\leftarrow$ $\phi$.
	\State  $Sort(\hat{\mathcal{S}})$ \Comment Sorting based on $\frac{\boldsymbol{\mathcal{D}}_{i}(\varUpsilon_{i-1})}{\bar{c}_i}$ for all $s_i \in \hat{\mathcal{S}}$
	  \For{$each$ $s_i \in \hat{\mathcal{S}}$}
	    \If{ $\frac{\bar{c}_i}{\boldsymbol{\mathcal{B}}'} \leq \left( \frac{\boldsymbol{\mathcal{D}}_{i}(\varUpsilon_{i-1})}{\boldsymbol{\mathcal{D}}_{i}(\varUpsilon_{i-1}) + \boldsymbol{\mathcal{D}}(\varUpsilon_{i-1})}\right)$}
	       \State $\varUpsilon$ $\leftarrow$ $\varUpsilon$ $\cup$ $\{s_{i}\}$
	     \EndIf
	     \EndFor
	      \Return $\varUpsilon$
 \end{algorithmic}
\end{algorithm}
\indent \textbf{\underline{Example} 1(c):} 
Considering the set-up shown in Figure \ref{fig:gull}. We have utilized the same cost vector as given in Figure \ref{fig:animals}. The patient's budget is given as 8. The quality vector is given as $\boldsymbol{\mathcal{Q}} = \{5,~1,~3,~5,~4,~5\}$. The set of nodes informed by the leaders $\{4,~3\}$ is given as $\{3,~5,~6,~4,~1,~2\}$. So, the nodes 3, 5, 6, 4, 1, and 2 are sorted based on quality contribution per cost and is given as: $\{1, 4, 6, 3, 2, 5\}$. First node 1 is considered and the condition $2 \leq 8 \cdotp (\frac{5}{5})$ for node 1 is satisfied. So, $\varUpsilon=\{1\}$. Next, node 4 will be considered and the condition $2 \leq 8 \cdotp (\frac{5}{10})$ for node 4 is satisfied. So, $\varUpsilon=\{1,~ 4\}$.
\paragraph{\textbf{Payment rule}}
The {\fontfamily{qcr}\selectfont Payment rule} is motivated by \cite{Singer:2014:BFM:2692359.2692366}. For each $s_i$, it is defined as the minimum of the doctor's proportional share and the threshold payment. $\ell$ is the largest index that satisfies the condition in line 2 of Algorithm 3.
  \begin{algorithm}[H]
\caption{TBC-DS Pricing Mechanism ($\varUpsilon$, $\boldsymbol{\mathcal{B}}'$, $\bar{\boldsymbol{\mathcal{C}}}$)}
\begin{algorithmic}[1]
      \Output $\hat{\boldsymbol{\mathcal{P}}}$ $\leftarrow$ $\phi$
	\For{$each$ $s_i$ $\in$ $\varUpsilon$}    
	       \State $\hat{\boldsymbol{\mathcal{P}}}_i \leftarrow min \left \{ \frac{\boldsymbol{\mathcal{D}}_{i}(\varUpsilon_{i-1}) \cdot \boldsymbol{\mathcal{B}}'}{\sum_{i \in \varUpsilon} \boldsymbol{\mathcal{D}}(\varUpsilon_{i})}, \frac{\boldsymbol{\mathcal{D}}_{i}(\varUpsilon_{i-1}) \cdot \bar{c}_{\ell+1}}{\boldsymbol{\mathcal{D}}_{\ell+1}(\varUpsilon_{\ell})} \right \}$; $\hat{\boldsymbol{\mathcal{P}}} \leftarrow \hat{\boldsymbol{\mathcal{P}}} \cup \hat{\boldsymbol{\mathcal{P}}}_i$    
	\EndFor
	      \Return $\hat{\boldsymbol{\mathcal{P}}}$       
\end{algorithmic}
\end{algorithm}
  \textbf{\underline{Example 1(d):}} The payment of doctors in $\varUpsilon=\{1,~ 4\}$ is: $\hat{\boldsymbol{\mathcal{P}}}_1$ = $min \left \{ \frac{5 \times 8}{10}, \frac{5 \times 4}{5} \right \} = 4$ and $\hat{\boldsymbol{\mathcal{P}}}_2$ = $min \left \{ \frac{5 \times 8}{10}, \frac{5 \times 4}{5} \right \} = 4$.   
 \begin{lemma}
The TBC mechanism is computationally efficient. 
\end{lemma}
\begin{proof}
In TBC-LI, line $1-5$ of Algorithm 3 is bounded above by \emph{m}. In Algorithm 4, for each iteration of \emph{for} loop line $3-14$ is bounded above by \emph{m}. As we have \emph{m} iterations in worst case, we have $O(m^2)$. Thus, the running time of TBC-LI mechanism is $O(m^2)$. In TBC-DS, line 1 of Algotrithm 5 takes $O(m \lg m)$ time. Line $2-6$ takes $O(m)$ time. So, overall running time of Algorithm 5 is $O(m \lg m) + O(m) = O(m \lg m)$. Line $1-3$ of Algorithm 6 takes time $O(m)$. Thus, running time of TBC-DS  mechanism is $O(m \lg m)$. The computational complexity of TBC mechanism is given as $O(m^2) + O(m \lg m) = O(m^2)$.       
\end{proof}
\section{Analysis of TBC-LI and TBC-DS}
\begin{lemma}
In TBC-LI, the total payment made to the doctors are within hospital's budget $\boldsymbol{\mathcal{B}}$.
\end{lemma}
\begin{proof}
The proof is motivated by \cite{Singer:2012:WFI:2124295.2124381}. As the maximum payment that any winning EC \emph{i} can be paid is $\boldsymbol{\Pi}_{i}^{k}$ = $\frac{\boldsymbol{\mathcal{B}} \cdot \boldsymbol{\mathcal{M}}_{\boldsymbol{\mathcal{C}}_k}(\varGamma_{i-1})}{\boldsymbol{\mathcal{I}}(\varGamma_{i-1} \cup \{i\})}$. The total payment of the ECs as leaders $i.e.$ $\mathcal{T}_{\mathcal{I}}^{\mathcal{P}}$ is given as:  
\begin{equation*}
\mathcal{T}_{\mathcal{I}}^{\mathcal{P}} = \sum_{i \in \hat{\mathcal{S}}} \mathcal{P}_i = \sum_{i=1}^{k} \boldsymbol{\mathcal{B}} \cdot \frac{\boldsymbol{\mathcal{M}}_{\boldsymbol{\mathcal{C}}_i}(\varGamma_{i-1})}{\boldsymbol{\mathcal{I}}(\varGamma_{i-1} \cup \{s_i\})} = \sum_{i=1}^{k} \boldsymbol{\mathcal{B}} \cdot \frac{\boldsymbol{\mathcal{M}}_{\boldsymbol{\mathcal{C}}_i}(\varGamma_{i-1})}{\boldsymbol{\mathcal{I}}(\varGamma_{i})}
\end{equation*}
\begin{equation*}
\leq \frac{\boldsymbol{\mathcal{B}}}{\mathcal{I}(\varGamma_k)} \cdot \sum_{i=1}^{k} \boldsymbol{\mathcal{M}}_{\boldsymbol{\mathcal{C}}_i}(\varGamma_{i-1}) = \frac{\boldsymbol{\mathcal{B}}}{\boldsymbol{\mathcal{I}}(\varGamma_k)}  \cdot \sum_{i=1}^{k} \underbrace{\boldsymbol{\mathcal{I}}(\varGamma_{i-1} \cup \{s_i\})}_{\substack{\text{Informed ECs}\\\text{by set $\varGamma_{i}$}}} - \underbrace{\boldsymbol{\mathcal{I}}(\varGamma_{i-1})}_{\substack{\text{Informed ECs}\\\text{ by set $\varGamma_{i-1}$}}}
\end{equation*}
\begin{equation*}
= \frac{\boldsymbol{\mathcal{B}}}{\boldsymbol{\mathcal{I}}(\varGamma_k)}  \cdot \sum_{i=1}^{k} \underbrace{\boldsymbol{\mathcal{I}}(\varGamma_{i})}_{\substack{\text{Informed}\\\text{ECs by}\\\text{set $\varGamma_{i}$}}} - \underbrace{\boldsymbol{\mathcal{I}}(\varGamma_{i-1})}_{\substack{\text{Informed ECs}\\\text{ by set $\varGamma_{i-1}$}}} \Rightarrow \mathcal{T}_{\mathcal{I}}^{\mathcal{P}} \leq \boldsymbol{\mathcal{B}}	
\end{equation*}
Hence, it is proved that the incentive compatible total payment do not exceed the budget.
 \end{proof}
\begin{lemma}
In TBC-LI mechanism, if any doctor $s_{i}$ comes ahead of its current position say $i'< i$ by declaring a cost $c_{i'} < c_{i}$ then,
\begin{equation*}
\small
\frac{\boldsymbol{\mathcal{B}}}{2}\left(\frac{\boldsymbol{\mathcal{M}}_{\boldsymbol{\mathcal{C}}_{i'}}(\varGamma_{i'-1})}{\boldsymbol{\mathcal{M}}_{\boldsymbol{\mathcal{C}}_{i'}}(\varGamma_{i'-1}) + \boldsymbol{\mathcal{I}}(\varGamma_{i'-1})}\right) > \frac{\boldsymbol{\mathcal{B}}}{2}\left(\frac{\boldsymbol{\mathcal{M}}_{\boldsymbol{\mathcal{C}}_{i}}(\varGamma_{i-1})}{\boldsymbol{\mathcal{M}}_{\boldsymbol{\mathcal{C}}_{i}}(\varGamma_{i-1}) + \mathcal{I}(\varGamma_{i-1})}\right)
\end{equation*}
\end{lemma}
\begin{proof}
If the EC \emph{i} by reporting $c_{i'}$ moves at position $i'$ such that $i' < i$ as depicted in \emph{Figure} 4 below:
\vspace*{5mm}
\begin{figure}[H]
\setlength{\unitlength}{4cm}
\begin{picture}(0,0)
\thicklines
\put(0.5,0){\line(1,0){1}}
\put(1.525,0){\circle{0.05}}
\put(0.477,0){\circle{0.05}}
\end{picture} 
\setlength{\unitlength}{0.3cm}
\begin{picture}(0,0)
\thinlines
\put(10.5,1.2){\vector(0,-1){1.0}}
\put(8.5,1.2){\vector(0,-1){1.0}}
\put(15.15,1.2){\vector(0,-1){1.0}}
\put(17.15,1.2){\vector(0,-1){1.0}}
\end{picture} 
\setlength{\unitlength}{1cm}
\begin{picture}(0,0)
\thicklines
\put(4.9,0){\circle*{0.2}}
\put(4.3,0){\circle*{0.2}}
\put(2.9,0){\circle*{0.2}}
\put(2.3,0){\circle*{0.2}}
\put(4.85,-0.4){$i$}
\put(3.9,-0.4){\small{$i-1$}}
\put(2.85,-0.4){$i'$}
\put(1.8,-0.4){\small{$i'-1$}}
\put(4.85,0.42){$c_i$}
\put(3.9,0.42){\small{$c_{i-1}$}}
\put(2.85,0.42){$c_{i'}$}
\put(1.8,0.42){\small{$c_{i'-1}$}}
\end{picture}

\vspace*{5mm}\caption{Pictorial representation}
\end{figure}
From the definition of $\mathcal{I}(\cdot)$ we can say: $\mathcal{I}(\varGamma_{i'-1}) < \mathcal{I}(\varGamma_{i-1})$.
As the set $\varGamma_{i'-1}$ is smaller as compared to the set $\varGamma_{i-1}$, so from the definition of the monotone sub-modular marginal contribution property, it can be said:
\begin{equation}\label{eq:15}
\small{
\underbrace{\boldsymbol{\mathcal{M}}_{\boldsymbol{\mathcal{C}}_{i'}}(\varGamma_{i'-1})}_{\substack{\text{Marginal contribution}\\\text{of $i'$ given $\varGamma_{i'-1}$}}} > \underbrace{\boldsymbol{\mathcal{M}}_{\boldsymbol{\mathcal{C}}_{i}}(\varGamma_{i-1})}_{\substack{\text{Marginal contribution}\\\text{of $i$ given $\varGamma_{i-1}$}}}
}
\end{equation} 
The number of ECs leaders by the set $\varGamma_{i'}$ will be less than the number of ECs leaders by $\varGamma_{i}$. Mathematically,
\begin{equation*}
\small{
\underbrace{\boldsymbol{\mathcal{M}}_{\boldsymbol{\mathcal{C}}_{i'}}(\varGamma_{i'-1})}_{\substack{\text{Marginal contribution}\\\text{of $i'$ given $\varGamma_{i'-1}$}}} +  \mathcal{I}(\varGamma_{i'-1}) < \underbrace{\boldsymbol{\mathcal{M}}_{\boldsymbol{\mathcal{C}}_{i}}(\varGamma_{i-1})}_{\substack{\text{Marginal contribution}\\\text{of $i$ given $\varGamma_{i-1}$}}} + \mathcal{I}(\varGamma_{i-1})
}
\end{equation*}
\begin{equation}\label{eq:16}
\small{
\frac{1}{\boldsymbol{\mathcal{M}}_{\boldsymbol{\mathcal{C}}_{i'}}(\varGamma_{i'-1}) +  \mathcal{I}(\varGamma_{i'-1})} > \frac{1}{\boldsymbol{\mathcal{M}}_{\boldsymbol{\mathcal{C}}_{i}}(\varGamma_{i-1}) + \boldsymbol{\mathcal{I}}(\varGamma_{i-1})}
}
\end{equation}
Combining \emph{equation} \ref{eq:15}, \emph{equation} \ref{eq:16} and multiplying both side by $\frac{\boldsymbol{\mathcal{B}}}{2}$, we get
\begin{equation*}
\small{
\frac{\boldsymbol{\mathcal{B}}}{2}\left(\frac{\boldsymbol{\mathcal{M}}_{\boldsymbol{\mathcal{C}}_{i'}}(\varGamma_{i'-1})}{\boldsymbol{\mathcal{M}}_{\boldsymbol{\mathcal{C}}_{i'}}(\varGamma_{i'-1}) + \mathcal{I}(\varGamma_{i'-1})}\right) > \frac{\boldsymbol{\mathcal{B}}}{2}\left(\frac{\boldsymbol{\mathcal{M}}_{\boldsymbol{\mathcal{C}}_{i}}(\varGamma_{i-1})}{\boldsymbol{\mathcal{M}}_{\boldsymbol{\mathcal{C}}_{i}}(\varGamma_{i-1}) + \boldsymbol{\mathcal{I}}(\varGamma_{i-1})}\right)
}
\end{equation*}
Hence, it is proved.
\end{proof}
\begin{theorem}
TBC-LI mechanism is monotone.
\end{theorem}
\begin{proof}
Fix \emph{i}, $c_{-i}$, $c_i$, and $c_{i'}$. For mechanism {\fontfamily{qcr}\selectfont TBC-LI mechanism} to be \emph{monotone}, we need to show that, any winning EC \emph{i} with private cost $c_i$ will still be considered in the winning set of ECs when declaring $c_{i'}$ such that $c_{i'}  < c_i$ or any losing EC \emph{i} with private cost $c_i$ will still be considered in the losing set of ECs when declaring $c_{i'}$ such that $c_{i'}  > c_i$. The proof is divided into two cases.
\begin{component}[Case 1]
In this case, the $i^{th}$ winning EC deviates and reveals a cost of consultation $c_{i'}$ $<$ $c_i$.
  Again two cases can happen. If the EC \emph{i} shows a small deviation in his/her (henceforth his) cost $c_i$ $i.e.$ $c_{i'}$ such that $c_{i'} < c_{i}$ and the current position of the EC \emph{i} remains unchanged. In this situation, it can still be considered in the winning set. It is to be noted that, if the EC \emph{i} reports a large deviation in his cost $c_{i}$ $i.e.$ $c_{i'}$ such that $c_{i'} < c_i$, then in this case by definition:  
 \begin{equation*}
 \small{
\frac{\boldsymbol{\mathcal{M}}_{\boldsymbol{\mathcal{C}}_1}(\varGamma_{0})}{c_1} \geq \frac{\boldsymbol{\mathcal{M}}_{\boldsymbol{\mathcal{C}}_2}(\varGamma_1)}{c_2} \geq \frac{\boldsymbol{\mathcal{M}}_{\boldsymbol{\mathcal{C}}_3}(\varGamma_2)}{c_3} \geq \ldots \geq \frac{\boldsymbol{\mathcal{M}}_{\boldsymbol{\mathcal{C}}_n}(\varGamma_{n-1})}{c_n}
}
 \end{equation*}
EC \emph{i} will be placed some position ahead (say $i'$) of its current position say $i$ $i.e.$ $i'<i$. This scenario is depicted in \emph{Figure} 5 below.
\begin{figure}[H]
\setlength{\unitlength}{4cm}
\begin{picture}(0,0)
\thicklines
\put(0.5,0){\line(1,0){1}}
\put(1.525,0){\circle{0.05}}
\put(0.477,0){\circle{0.05}}
\end{picture} 
\setlength{\unitlength}{0.3cm}
\begin{picture}(0,0)
\thinlines
\put(10.5,1.2){\vector(0,-1){1.0}}
\put(8.5,1.2){\vector(0,-1){1.0}}
\put(15.15,1.2){\vector(0,-1){1.0}}
\put(17.15,1.2){\vector(0,-1){1.0}}
\end{picture} 
\setlength{\unitlength}{1cm}
\begin{picture}(0,0)
\thicklines
\put(4.9,0){\circle*{0.2}}
\put(4.3,0){\circle*{0.2}}
\put(2.9,0){\circle*{0.2}}
\put(2.3,0){\circle*{0.2}}
\put(4.85,-0.4){$i$}
\put(3.9,-0.4){\small{$i-1$}}
\put(2.85,-0.4){$i'$}
\put(1.8,-0.4){\small{$i'-1$}}
\put(4.85,0.42){$c_i$}
\put(3.9,0.42){\small{$c_{i-1}$}}
\put(2.85,0.42){$c_{i'}$}
\put(1.8,0.42){\small{$c_{i'-1}$}}
\end{picture}
\vspace*{5mm}\caption{Pictorial representation}
\end{figure}
From \emph{Lemma} 6 it can be said that if EC $i$ is placed some position ahead by revealing a cost $c_{i'} < c_{i}$ then it must satisfy 
\begin{equation}\label{eq:17}
\small{
\frac{\boldsymbol{\mathcal{M}}_{\boldsymbol{\mathcal{C}}_{i'}}(\varGamma_{i'-1})}{\boldsymbol{\mathcal{M}}_{\boldsymbol{\mathcal{C}}_{i'}}(\varGamma_{i'-1}) + \boldsymbol{\mathcal{I}}(\varGamma_{i'-1})} > \frac{\boldsymbol{\mathcal{M}}_{\boldsymbol{\mathcal{C}}_{i}}(\varGamma_{i-1})}{\boldsymbol{\mathcal{M}}_{\boldsymbol{\mathcal{C}}_{i}}(\varGamma_{i-1}) + \boldsymbol{\mathcal{I}}(\varGamma_{i-1})}
}
\end{equation} 
Let us suppose for the sake of contradiction that, when the EC $i \in \mathcal{S}$ comes ahead in ordering say at some position $i'$ such that $c_{i'} < c_i$, then it is not considered in the winning set of the EC because it is not satisfying the given budget. If this is the case, then it means that:
\begin{equation}\label{eq:18}
\small{
c_{i'} > \frac{\boldsymbol{\mathcal{B}}}{2}\left(\frac{\boldsymbol{\mathcal{M}}_{\boldsymbol{\mathcal{C}}_{i'}}(\varGamma_{i'-1})}{\boldsymbol{\mathcal{M}}_{\boldsymbol{\mathcal{C}}_{i'}}(\varGamma_{i'-1}) + \boldsymbol{\mathcal{I}}(\varGamma_{i'-1})}\right)
} 
\end{equation} 
Combining our assumption $c_{i'} < c_{i}$ and \emph{equation} \ref{eq:18} it can be concluded that:
\begin{equation}\label{eq:19}
\small{
 \frac{\boldsymbol{\mathcal{B}}}{2}\left(\frac{\boldsymbol{\mathcal{M}}_{\boldsymbol{\mathcal{C}}_{i'}}(\varGamma_{i'-1})}{\boldsymbol{\mathcal{M}}_{\boldsymbol{\mathcal{C}}_{i'}}(\varGamma_{i'-1}) + \mathcal{I}(\varGamma_{i'-1})}\right) < c_{i'} < c_i
 }
\end{equation}
Using condition in line 2 of \emph{Algorithm 3} and \emph{equation} \ref{eq:19}, we can say that:

\small{
\begin{equation*}
 \frac{\boldsymbol{\mathcal{B}}}{2}\left(\frac{\boldsymbol{\mathcal{M}}_{\boldsymbol{\mathcal{C}}_{i'}}(\varGamma_{i'-1})}{\boldsymbol{\mathcal{M}}_{\boldsymbol{\mathcal{C}}_{i'}}(\varGamma_{i'-1}) + \mathcal{I}(\varGamma_{i'-1})}\right) < c_{i'} < c_i < \frac{\boldsymbol{\mathcal{B}}}{2}\left(\frac{\boldsymbol{\mathcal{M}}_{\boldsymbol{\mathcal{C}}_{i}}(\varGamma_{i-1})}{\boldsymbol{\mathcal{M}}_{\boldsymbol{\mathcal{C}}_{i}}(\varGamma_{i-1}) + \mathcal{I}(\varGamma_{i-1})}\right)
\end{equation*}
}
\begin{equation*}
\small{
\hspace*{47mm} 
}
\end{equation*}
\begin{equation*}
\small{
\Rightarrow \frac{\boldsymbol{\mathcal{B}}}{2}\left(\frac{\boldsymbol{\mathcal{M}}_{\boldsymbol{\mathcal{C}}_{i'}}(\varGamma_{i'-1})}{\boldsymbol{\mathcal{M}}_{\boldsymbol{\mathcal{C}}_{i'}}(\varGamma_{i'-1}) + \mathcal{I}(\varGamma_{i'-1})}\right)< \frac{\boldsymbol{\mathcal{B}}}{2}\left(\frac{\boldsymbol{\mathcal{M}}_{\boldsymbol{\mathcal{C}}_{i}}(\varGamma_{i-1})}{\boldsymbol{\mathcal{M}}_{\boldsymbol{\mathcal{C}}_{i}}(\varGamma_{i-1}) + \mathcal{I}(\varGamma_{i-1})}\right)
}
\end{equation*}
\begin{equation*}
\small{
\Rightarrow \frac{\boldsymbol{\mathcal{M}}_{\boldsymbol{\mathcal{C}}_{i'}}(\varGamma_{i'-1})}{\boldsymbol{\mathcal{M}}_{\boldsymbol{\mathcal{C}}_{i'}}(\varGamma_{i'-1}) + \boldsymbol{\mathcal{I}}(\varGamma_{i'-1})}< \frac{\boldsymbol{\mathcal{M}}_{\boldsymbol{\mathcal{C}}_{i}}(\varGamma_{i-1})}{\boldsymbol{\mathcal{M}}_{\boldsymbol{\mathcal{C}}_{i}}(\varGamma_{i-1}) + \boldsymbol{\mathcal{I}}(\varGamma_{i-1})}
}
\end{equation*}
So, it is a contradiction.
\end{component}
\begin{component}[Case 2]
In this case, the $i^{th}$ losing EC deviates and reveals a cost of consultation $c_{i'}$ $>$ $c_i$.
  Again two cases can happen. If the EC \emph{i} shows a small deviation in his/her (henceforth his) cost $c_i$ $i.e.$ $c_{i'}$ such that $c_{i'} > c_{i}$ and the current position of the EC \emph{i} remains unchanged. In this situation, it will be considered in the losing set. It is to be noted that, if the EC \emph{i} reports a large deviation in his cost $c_{i}$ $i.e.$ $c_{i'}$ such that $c_{i'} > c_i$, then in this case by definition:  
 \begin{equation*}
\frac{\boldsymbol{\mathcal{M}}_{\boldsymbol{\mathcal{C}}_1}(\varGamma_{0})}{c_1} \geq \frac{\boldsymbol{\mathcal{M}}_{\boldsymbol{\mathcal{C}}_2}(\varGamma_1)}{c_2} \geq \frac{\boldsymbol{\mathcal{M}}_{\boldsymbol{\mathcal{C}}_3}(\varGamma_2)}{c_3} \geq \ldots \geq \frac{\boldsymbol{\mathcal{M}}_{\boldsymbol{\mathcal{C}}_n}(\varGamma_{n-1})}{c_n}
 \end{equation*}
EC \emph{i} will be placed some position ahead (say $i'$) of its current position say $i$ $i.e.$ $i'>i$. This scenario is depicted in \emph{Figure} 6 below.
\begin{figure}[H]
\setlength{\unitlength}{4cm}
\begin{picture}(0,0)
\thicklines
\put(0.5,0){\line(1,0){1}}
\put(1.525,0){\circle{0.05}}
\put(0.477,0){\circle{0.05}}
\end{picture} 
\setlength{\unitlength}{0.3cm}
\begin{picture}(0,0)
\thinlines
\put(10.5,1.2){\vector(0,-1){1.0}}
\put(8.5,1.2){\vector(0,-1){1.0}}
\put(15.15,1.2){\vector(0,-1){1.0}}
\put(17.15,1.2){\vector(0,-1){1.0}}
\end{picture} 
\setlength{\unitlength}{1cm}
\begin{picture}(0,0)
\thicklines
\put(4.9,0){\circle*{0.2}}
\put(4.3,0){\circle*{0.2}}
\put(2.9,0){\circle*{0.2}}
\put(2.3,0){\circle*{0.2}}
\put(4.85,-0.4){$i$}
\put(3.9,-0.4){\small{$i'-1$}}
\put(2.85,-0.4){$i$}
\put(1.8,-0.4){\small{$i-1$}}
\put(4.85,0.42){$c_{i'}$}
\put(3.9,0.42){\small{$c_{i'-1}$}}
\put(2.85,0.42){$c_{i}$}
\put(1.8,0.42){\small{$c_{i-1}$}}
\end{picture}
\vspace*{5mm}\caption{Pictorial representation}
\end{figure}
Analogous to the statement given in \emph{Lemma} 6 it can be said that if EC $i$ is placed some position ahead by revealing a cost $c_{i'} > c_{i}$ then it must satisfy 
\begin{equation}\label{eq:20}
\frac{\boldsymbol{\mathcal{M}}_{\boldsymbol{\mathcal{C}}_{i'}}(\varGamma_{i'-1})}{\boldsymbol{\mathcal{M}}_{\boldsymbol{\mathcal{C}}_{i'}}(\varGamma_{i'-1}) + \boldsymbol{\mathcal{I}}(\varGamma_{i'-1})} < \frac{\boldsymbol{\mathcal{M}}_{\boldsymbol{\mathcal{C}}_{i}}(\varGamma_{i-1})}{\boldsymbol{\mathcal{M}}_{\boldsymbol{\mathcal{C}}_{i}}(\varGamma_{i-1}) + \boldsymbol{\mathcal{I}}(\varGamma_{i-1})}
\end{equation}
Let us suppose for the sake of contradiction that, when the EC $i \in \mathcal{S}$ comes ahead in ordering say at some position $i'$ such that $c_i' > c_i$, then it is not considered in the losing set of the EC because it is satisfying the given budget. If this is the case, then it means that:
\begin{equation}\label{eq:21}
c_{i'} < \frac{\boldsymbol{\mathcal{B}}}{2}\left(\frac{\boldsymbol{\mathcal{M}}_{\boldsymbol{\mathcal{C}}_{i'}}(\varGamma_{i'-1})}{\boldsymbol{\mathcal{M}}_{\boldsymbol{\mathcal{C}}_{i'}}(\varGamma_{i'-1}) + \mathcal{I}(\varGamma_{i'-1})}\right) 
\end{equation} 
Combining our assumption $c_{i'} >c_{i}$ and \emph{equation} \ref{eq:21} it can be concluded that:
\begin{equation}\label{eq:22}
 \frac{\boldsymbol{\mathcal{B}}}{2}\left(\frac{\boldsymbol{\mathcal{M}}_{\boldsymbol{\mathcal{C}}_{i'}}(\varGamma_{i'-1})}{\boldsymbol{\mathcal{M}}_{\boldsymbol{\mathcal{C}}_{i'}}(\varGamma_{i'-1}) + \mathcal{I}(\varGamma_{i'-1})}\right) > c_{i'} > c_i
\end{equation}
Using condition in line 2 of \emph{Algorithm 3} and \emph{equation} \ref{eq:22}, we can say that:
\begin{equation*}
 \frac{\boldsymbol{\mathcal{B}}}{2}\left(\frac{\boldsymbol{\mathcal{M}}_{\boldsymbol{\mathcal{C}}_{i'}}(\varGamma_{i'-1})}{\boldsymbol{\mathcal{M}}_{\boldsymbol{\mathcal{C}}_{i'}}(\varGamma_{i'-1}) + \mathcal{I}(\varGamma_{i'-1})}\right) > c_{i'} > c_i > \frac{\boldsymbol{\mathcal{B}}}{2}\left(\frac{\boldsymbol{\mathcal{M}}_{\boldsymbol{\mathcal{C}}_{i}}(\varGamma_{i-1})}{\boldsymbol{\mathcal{M}}_{\boldsymbol{\mathcal{C}}_{i}}(\varGamma_{i-1}) + \mathcal{I}(\varGamma_{i-1})}\right)
\end{equation*}
\begin{equation*}
\small{
\hspace*{47mm} 
}
\end{equation*}
\begin{equation*}
\Rightarrow \frac{\boldsymbol{\mathcal{B}}}{2}\left(\frac{\boldsymbol{\mathcal{M}}_{\boldsymbol{\mathcal{C}}_{i'}}(\varGamma_{i'-1})}{\boldsymbol{\mathcal{M}}_{\boldsymbol{\mathcal{C}}_{i'}}(\varGamma_{i'-1}) + \mathcal{I}(\varGamma_{i'-1})}\right)> \frac{\boldsymbol{\mathcal{B}}}{2}\left(\frac{\boldsymbol{\mathcal{M}}_{\boldsymbol{\mathcal{C}}_{i}}(\varGamma_{i-1})}{\boldsymbol{\mathcal{M}}_{\boldsymbol{\mathcal{C}}_{i}}(\varGamma_{i-1}) + \mathcal{I}(\varGamma_{i-1})}\right)
\end{equation*}
\begin{equation*}
\Rightarrow \frac{\boldsymbol{\mathcal{M}}_{\boldsymbol{\mathcal{C}}_{i'}}(\varGamma_{i'-1})}{\boldsymbol{\mathcal{M}}_{\boldsymbol{\mathcal{C}}_{i'}}(\varGamma_{i'-1}) + \boldsymbol{\mathcal{I}}(\varGamma_{i'-1})}> \frac{\boldsymbol{\mathcal{M}}_{\boldsymbol{\mathcal{C}}_{i}}(\varGamma_{i-1})}{\boldsymbol{\mathcal{M}}_{\boldsymbol{\mathcal{C}}_{i}}(\varGamma_{i-1}) + \boldsymbol{\mathcal{I}}(\varGamma_{i-1})}
\end{equation*}
So, it is a contradiction.
\end{component}
Hence, the theorem is proved.
 \end{proof}
\begin{theorem}
In TBC-DS, the function $\boldsymbol{\mathcal{D}}: 2^{\mathcal{S}} \Rightarrow \mathcal{R}_{\geq 0}$ is: 
\begin{itemize}
\item[(a)] \textbf{\emph{monotone}:} If $\mathcal{S} \subseteq \mathcal{F}$ then $\boldsymbol{\mathcal{D}} (\mathcal{S}) \leq \boldsymbol{\mathcal{D}} (\mathcal{F}) $, and
\item[(b)] \textbf{\emph{submodular}:} If $\boldsymbol{\mathcal{D}} (\mathcal{S} \cup \{i\}) - \boldsymbol{\mathcal{D}} (\mathcal{S}) \geq \boldsymbol{\mathcal{D}} (\mathcal{F} \cup \{i\}) -\boldsymbol{\mathcal{D}} (\mathcal{F})$ $\forall \mathcal{S} \subseteq \mathcal{F}$. 
\end{itemize}
\end{theorem}
\begin{proof}
To prove that the function is indeed monotone, let us suppose for the sake of contradiction that, if $\mathcal{S} \subseteq \mathcal{F}$ then $\boldsymbol{\mathcal{D}} (\mathcal{S}) > \boldsymbol{\mathcal{D}} (\mathcal{F})$.
From the definition of \emph{quality} function, we can say $\boldsymbol{\mathcal{D}} (\mathcal{S}) > \boldsymbol{\mathcal{D}} (\mathcal{F})$=$\sum_{i \in \mathcal{S}} \mathcal{Q}_i > \sum_{i \in \mathcal{F}} \mathcal{Q}_i$.
It is to be noted that under the given condition $\mathcal{S} \subseteq \mathcal{F}$, the sum of all the $\mathcal{Q}_{i}'s$ over the set $\mathcal{F}$ will be greater than the sum of all the $\mathcal{Q}_{i}'s$ over the set $\mathcal{S}$ $i.e.$ $\sum_{i \in \mathcal{S}} \mathcal{Q}_i \leq \sum_{i \in \mathcal{F}} \mathcal{Q}_i$. So, the inequality $\boldsymbol{\mathcal{D}} (\mathcal{S}) > \boldsymbol{\mathcal{D}} (\mathcal{F})$ = $\sum_{i \in \mathcal{S}} \mathcal{Q}_i > \sum_{i \in \mathcal{F}} \mathcal{Q}_i$ cannot be true. Our assumption contradicts. Hence, the inequality if $\mathcal{S} \subseteq \mathcal{F}$ then $\boldsymbol{\mathcal{D}} (\mathcal{S}) \leq \boldsymbol{\mathcal{D}} (\mathcal{F}) $ holds and the given function is \emph{monotone}.\\
To prove that the function is \emph{submodular}, as a thought experiment one can say that adding the same quantity $i.e.$ in this case the quality value of any agent $s_i$ to the sets having relation $\mathcal{S} \subseteq \mathcal{F}$ will reflect the contribution by any agent $s_i$ more in $\mathcal{S}$ than in $\mathcal{F}$. This completes the proof. 	
 \end{proof}   
\begin{theorem}
TBC-LI is truthful.
\end{theorem}
\begin{proof}
Fix EC \emph{j}, $c_{-j}$, $c_j$, and $c_{j'}$. For {\fontfamily{qcr}\selectfont TBC-LI mechanism} to be \emph{truthful}, we need to show that, it is not beneficial for any EC \emph{j} to \emph{underbid} or \emph{overbid} say $c_{j'}$ such that $c_{j'}<c_j$ or $c_{j'} > c_{j}$ respectively. For each of the above possible scenarios, the proof is divided into two cases. Before going into the different cases let's consider the case where the EC \emph{j} is reporting his true cost $c_j$. The pictorial representation of the possible set-up with \emph{n} ECs are shown in Figure \ref{theo4:4}. 
\begin{figure}[H]
\centering
\includegraphics[scale=0.6]{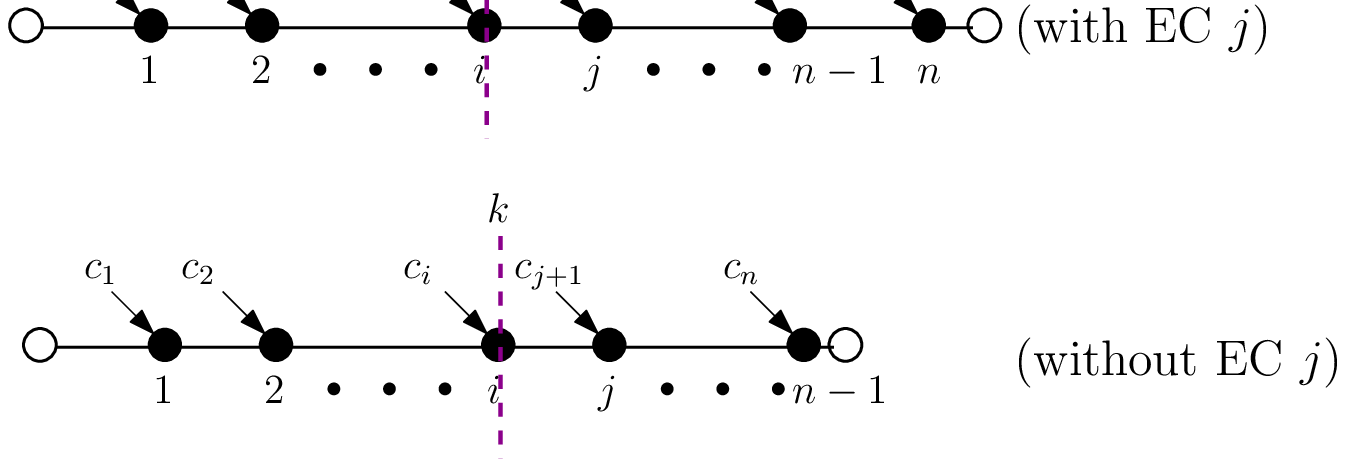}
\caption{Pictorial representation}
\label{theo4:4}
\end{figure} 
The values from 1 to \emph{n} represents the position (or index). Currently, our analysis lies around the index $k$ and $k+1$; where $k$ denote the index of the last EC $\ell$ that respects the allocation condition given in line 2 of \emph{Algorithm 3}. As the ECs are sorted based on the marginal contribution per cost, so we can write 
\begin{equation}\label{eq:23}
\frac{\boldsymbol{\mathcal{M}}_{\boldsymbol{\mathcal{C}}_i}(\varGamma_{i-1})}{c_i} \geq \frac{\boldsymbol{\mathcal{M}}_{\boldsymbol{\mathcal{C}}_{j}}(\varGamma_{i-1})}{c_j} \Rightarrow c_j \geq \frac{\boldsymbol{\mathcal{M}}_{\boldsymbol{\mathcal{C}}_j}(\varGamma_{i-1}) \cdot c_i}{\boldsymbol{\mathcal{M}}_{\boldsymbol{\mathcal{C}}_{i}}(\varGamma_{i-1})}
\end{equation} 
By using line 12 of \emph{algorithm 4} and equation \ref{eq:23} it can be easily seen that, $c_j \geq \boldsymbol{\mathcal{C}}_{j}^{i}$. If this is the case then we can say $c_j \geq \boldsymbol{\Pi}_{j}^{i}$. In order to be allocated EC \emph{j} must satisfy $\boldsymbol{\mathcal{C}}_{j}^{i} \leq \boldsymbol{\Pi}_{j}^{i}$ otherwise $\boldsymbol{\mathcal{C}}_{j}^{i} > \boldsymbol{\Pi}_{j}^{i}$ means not allocated. As we are taking the payment as:
\begin{itemize}
\item If $\boldsymbol{\mathcal{C}}_{j}^{i} = \boldsymbol{\Pi}_{j}^{i}$ $\Rightarrow$ $\boldsymbol{\mathcal{P}}_{\Gamma_j} = \min\{\boldsymbol{\mathcal{C}}_{j}^{i},\boldsymbol{\Pi}_{j}^{i} \}$ = $\boldsymbol{\Pi}_{j}^{i}$ = $\boldsymbol{\mathcal{C}}_{j}^{i}$ $\leq$ $c_j$
\item  If $\boldsymbol{\mathcal{C}}_{j}^{i} >\boldsymbol{\Pi}_{j}^{i}$ $\Rightarrow$ $\boldsymbol{\mathcal{P}}_{\Gamma_j} = \min\{\boldsymbol{\mathcal{C}}_{j}^{i},\boldsymbol{\Pi}_{j}^{i} \}$ = $\boldsymbol{\Pi}_{j}^{i}$ = $\boldsymbol{\Pi}_{j}^{i}$ $\leq$ $c_j$
\item If $\boldsymbol{\mathcal{C}}_{j}^{i} < \boldsymbol{\Pi}_{j}^{i}$ $\Rightarrow$ $\boldsymbol{\mathcal{P}}_{\Gamma_j} = \min\{\boldsymbol{\mathcal{C}}_{j}^{i},\boldsymbol{\Pi}_{j}^{i} \}$ = $\boldsymbol{\mathcal{C}}_{j}^{i}$ $\leq$ $c_j$
\end{itemize} 
If this is the case, then it can be concluded that $\boldsymbol{\mathcal{C}}_{j}^{i}  \leq  c_j$ or $\boldsymbol{\Pi}_{j}^{i} \leq c_j$. As the payment is less than the actual cost. Hence not allocated. Coming back to our \emph{underbid} and \emph{overbid} cases.

\paragraph*{\textbf{Scenario 1:} \textbf{Underbidding ($c_{j'} < c_j$)}}
In this case, the $j^{th}$ EC deviates and reveals a cost of consultation $c_{j'}$ $<$ $c_j$. This scenario give rise to two cases.
\paragraph*{\textbf{Case 1:}\textbf{When EC \emph{j} is in losing set.}}
If the EC \emph{j} shows a small deviation in his/her (henceforth his) cost $i.e.$ $c_{j'}$ such 
\begin{figure}[H]
\centering
\includegraphics[scale=0.6]{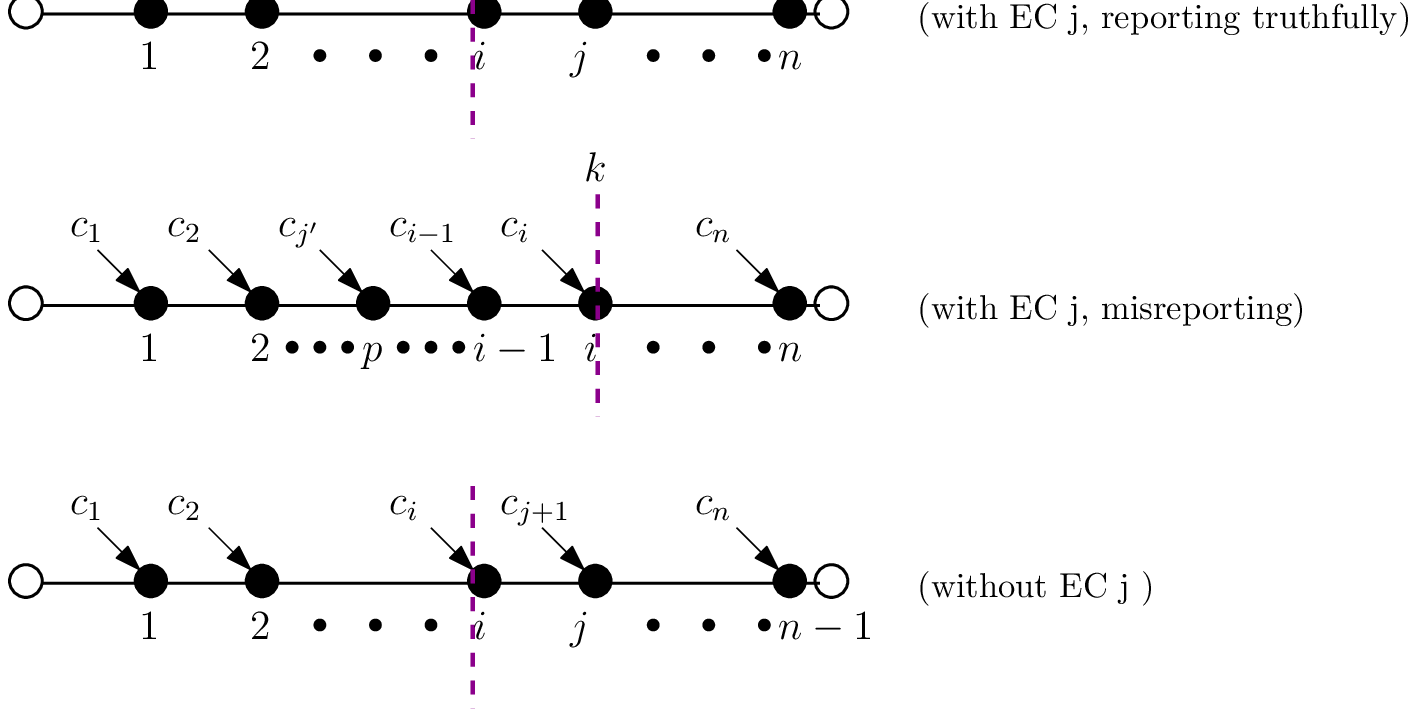}
\caption{Pictorial representation}
\label{theo5:5}
\end{figure}
that $c_{j'} < c_{j}$ and the current position of the EC \emph{j} remains unchanged. 
In this situation, it can still be considered in the losing set. It is to be noted that, if the EC \emph{j} reports a large deviation in his cost $c_{j}$ $i.e.$ $c_{j'}$ in this case it will belong to winning set and will appear before EC \emph{i} as shown in Figure \ref{theo5:5}.
 As the ECs are sorted based on the marginal contribution per cost, so we can write: 
\begin{equation*}
\frac{\boldsymbol{\mathcal{M}}_{\boldsymbol{\mathcal{C}}_j}(\varGamma_{i-1})}{c_{j'}} \geq \frac{\boldsymbol{\mathcal{M}}_{\boldsymbol{\mathcal{C}}_{i}}(\varGamma_{i-1})}{c_i} \Rightarrow c_{j'} \leq \frac{\boldsymbol{\mathcal{M}}_{\boldsymbol{\mathcal{C}}_j}(\varGamma_{i-1}) \cdot c_i}{\boldsymbol{\mathcal{M}}_{\boldsymbol{\mathcal{C}}_{i}}(\varGamma_{i-1})}
\end{equation*}
and $c_{j'} \leq \frac{\boldsymbol{\mathcal{M}}_{\boldsymbol{\mathcal{C}}_j}(\varGamma_{i-1}) \cdot c_i}{\boldsymbol{\mathcal{M}}_{\boldsymbol{\mathcal{C}}_{i}}(\varGamma_{i-1})} = \boldsymbol{\mathcal{C}}_{j}^{i}$ because from above we have got the relation $\boldsymbol{\mathcal{C}}_{j}^{i} \leq c_j$. This will lead to $c_{j'} \leq \boldsymbol{\mathcal{C}}_{j}^{i} \leq c_j$. The EC \emph{j} is paid less than the actual cost. 
\paragraph*{\textbf{Case 2:} \textbf{When EC \emph{j} is in winning set.}}
If the EC \emph{j} shows a deviation in his cost such that $c_{j'}<c_j$ it will still belong to winning set and will appear before EC \emph{i} as shown in Figure \ref{theo6:6}. As the ECs are sorted based on the marginal contribution per cost, so we can write: 
\begin{equation*}
\frac{\boldsymbol{\mathcal{M}}_{\boldsymbol{\mathcal{C}}_j}(\varGamma_{l-1})}{c_{j}} \geq \frac{\boldsymbol{\mathcal{M}}_{\boldsymbol{\mathcal{C}}_{i}}(\varGamma_{i-1})}{c_i} \Rightarrow c_{j} \leq \frac{\boldsymbol{\mathcal{M}}_{\boldsymbol{\mathcal{C}}_j}(\varGamma_{l-1}) \cdot c_i}{\boldsymbol{\mathcal{M}}_{\boldsymbol{\mathcal{C}}_{i}}(\varGamma_{i-1})} = \boldsymbol{\mathcal{C}}_j^l
\end{equation*}
and for the case when the EC \emph{j} deviates, then
 \begin{equation*}
\frac{\boldsymbol{\mathcal{M}}_{\boldsymbol{\mathcal{C}}_j}(\varGamma_{t-1})}{c_{j'}} \geq \frac{\boldsymbol{\mathcal{M}}_{\boldsymbol{\mathcal{C}}_{i}}(\varGamma_{i-1})}{c_i} \Rightarrow c_{j'} \leq \frac{\boldsymbol{\mathcal{M}}_{\boldsymbol{\mathcal{C}}_j}(\varGamma_{t-1}) \cdot c_i}{\boldsymbol{\mathcal{M}}_{\boldsymbol{\mathcal{C}}_{i}}(\varGamma_{i-1})} = \boldsymbol{\mathcal{C}}_j^l
\end{equation*} 
From above two equations it is clear that no matter what cost EC \emph{j} is bidding, he will still be winning and be paid an amount $\boldsymbol{\mathcal{C}}_{j}^{l}$.
 Hence, considering \emph{Case 1} and \emph{Case 2} it can be concluded that EC $j$ does not gain by underbidding there true cost. In similar fashion, we can write the above mentioned equation for any position \emph{i} before \emph{k} and in the same way we can prove that $\boldsymbol{\mathcal{C}}_{j}^{i} \leq c_j$.
  \begin{figure}[H]
\centering
\includegraphics[scale=0.55]{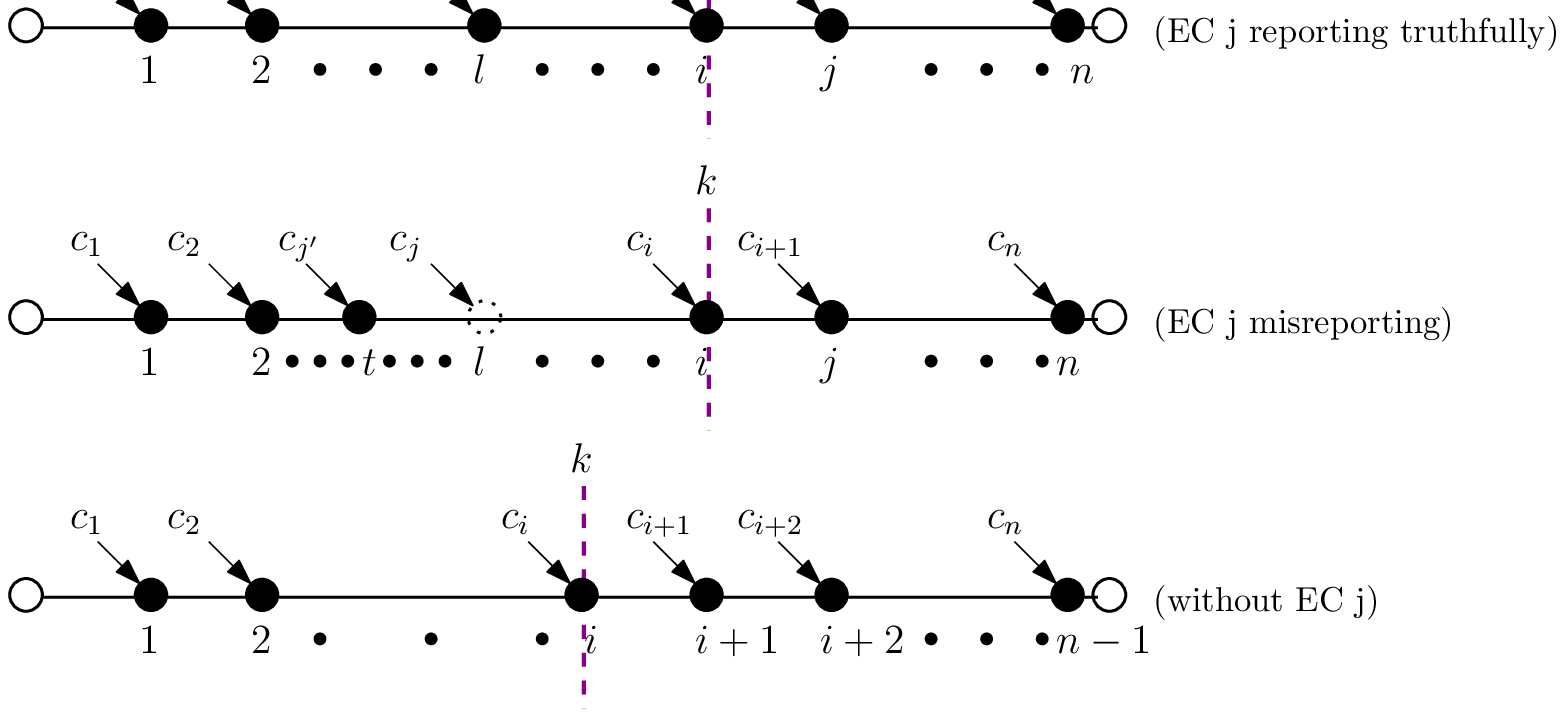}
\caption{Pictorial representation}
\label{theo6:6}
\end{figure}
 

\paragraph*{\textbf{Scenario 2:} \textbf{Overbidding ($c_{j'} > c_j$)}}
In this case, the $j^{th}$ EC deviates and reveals a cost of consultation $c_{j'}$ $>$ $c_j$. This scenario gives rise to two cases.
\paragraph*{\textbf{Case 1:} \textbf{When EC \emph{j} is in losing set.}}
If the EC \emph{j} shows a small deviation in his/her (henceforth his) cost $i.e.$ $c_{j'}$ such that $c_{j'} > c_{j}$ and the current position of the EC \emph{j} remains unchanged. In this situation, it can still be considered in the losing set.
\begin{figure}[H]
\centering
\includegraphics[scale=0.55]{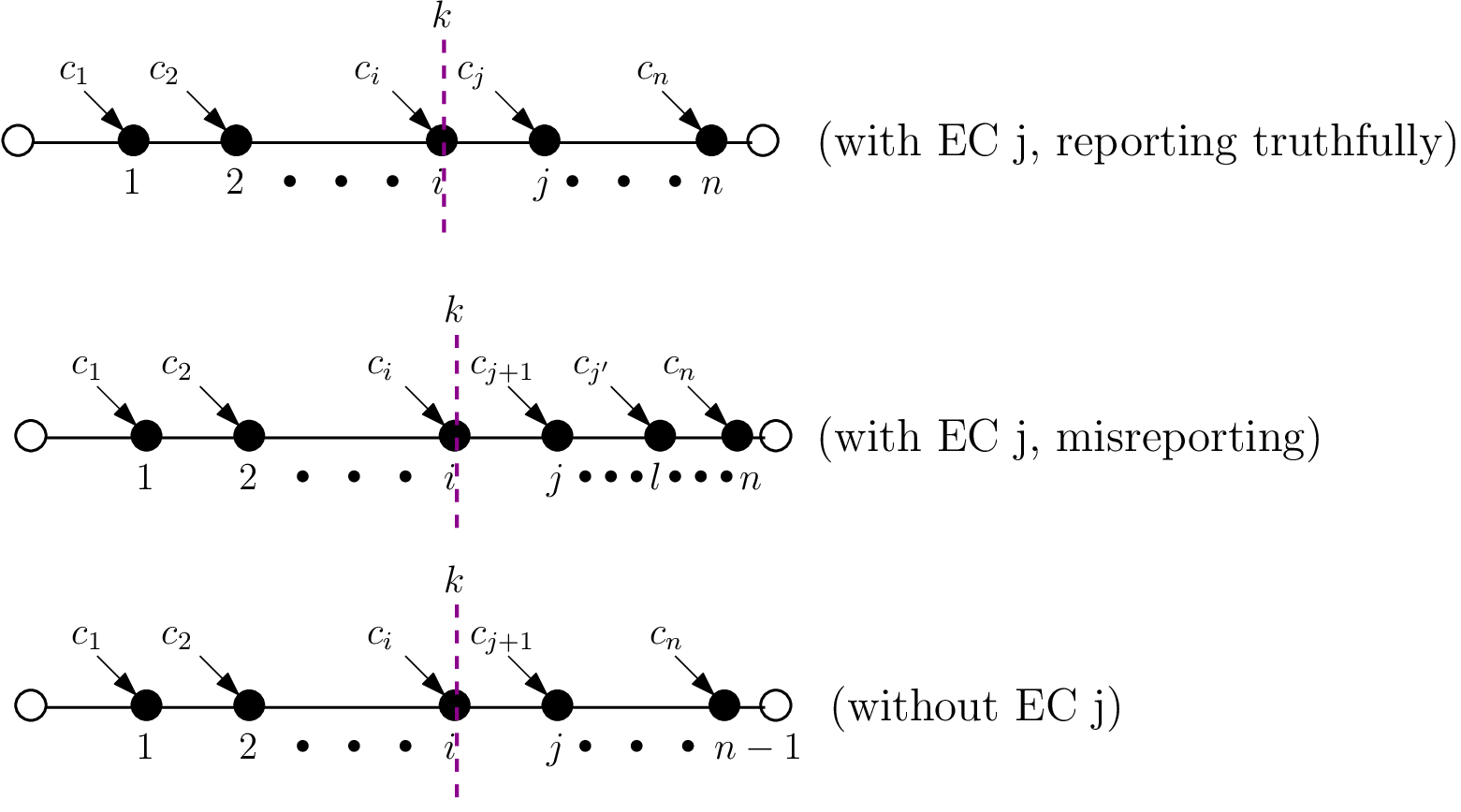}
\caption{Pictorial representation}
\label{theo7:7}
\end{figure}
 It is to be noted that, if the EC \emph{j} reports a large deviation in his cost $c_{j}$ $i.e.$ $c_{j'}$ in this case it will belong to losing set and will still appear after EC \emph{i} as shown in Figure \ref{theo7:7}. As the ECs are sorted based on the marginal contribution per cost, so we can write:
\begin{equation*}
\frac{\boldsymbol{\mathcal{M}}_{\boldsymbol{\mathcal{C}}_i}(\varGamma_{i-1})}{c_{i}} \geq \frac{\boldsymbol{\mathcal{M}}_{\boldsymbol{\mathcal{C}}_{j}}(\varGamma_{i-1})}{c_i} \Rightarrow c_{j'} \leq \frac{\boldsymbol{\mathcal{M}}_{\boldsymbol{\mathcal{C}}_j}(\varGamma_{i-1}) \cdot c_i}{\boldsymbol{\mathcal{M}}_{\boldsymbol{\mathcal{C}}_{i}}(\varGamma_{i-1})}
\end{equation*}
and $c_{j'} \leq \frac{\boldsymbol{\mathcal{M}}_{\boldsymbol{\mathcal{C}}_j}(\varGamma_{i-1}) \cdot c_i}{\boldsymbol{\mathcal{M}}_{\boldsymbol{\mathcal{C}}_{i}}(\varGamma_{i-1})} = \boldsymbol{\mathcal{C}}_{j}^{i}$. 
 From above we have got the relation $\boldsymbol{\mathcal{C}}_{j}^{i} \leq c_j$. This will lead to $c_{j'} \leq \boldsymbol{\mathcal{C}}_{j}^{i} \leq c_j$. The EC \emph{j} is paid less than the actual cost. 
\paragraph*{\textbf{Case 2:} \textbf{When EC \emph{j} is in winning set.}}
If the EC \emph{j} shows a deviation in his cost such that $c_{j'}>c_j$ and the current position of the EC \emph{j} remains unchanged. In this situation, it can still be considered in the winning set. It is to be noted that, if the EC \emph{j} reports a large deviation in his cost $c_{j}$ $i.e.$ $c_{j'}$ in this case it will belong to losing set and will appear after EC \emph{i} as shown in Figure \ref{theo8:8}.
Utilizing the definition of the marginal contribution per cost sorting and figure above:
\begin{equation*}
\frac{\boldsymbol{\mathcal{M}}_{\boldsymbol{\mathcal{C}}_j}(\varGamma_{l-1})}{c_{j}} \geq \frac{\boldsymbol{\mathcal{M}}_{\boldsymbol{\mathcal{C}}_{i}}(\varGamma_{i-1})}{c_i} \Rightarrow c_{j} \leq \frac{\boldsymbol{\mathcal{M}}_{\boldsymbol{\mathcal{C}}_j}(\varGamma_{l-1}) \cdot c_i}{\boldsymbol{\mathcal{M}}_{\boldsymbol{\mathcal{C}}_{i}}(\varGamma_{i-1})} = \boldsymbol{\mathcal{C}}_j^{l}
\end{equation*}
and for the case when the EC \emph{j} deviates, then
 \begin{equation*}
\frac{\boldsymbol{\mathcal{M}}_{\boldsymbol{\mathcal{C}}_i}(\varGamma_{i-1})}{c_{i}} \geq \frac{\boldsymbol{\mathcal{M}}_{\boldsymbol{\mathcal{C}}_{j}}(\varGamma_{t-1})}{c_{j'}} \Rightarrow c_{j'} \geq \frac{\boldsymbol{\mathcal{M}}_{\boldsymbol{\mathcal{C}}_j}(\varGamma_{t-1}) \cdot c_i}{\boldsymbol{\mathcal{M}}_{\boldsymbol{\mathcal{C}}_{i}}(\varGamma_{i-1})} = \boldsymbol{\mathcal{C}}_j^l
\end{equation*}
Now, if EC \emph{j} deviates by large amount then it will belong to the losing set. From above two equations it is clear that no matter what cost EC \emph{j} is bidding, he will be paid $\boldsymbol{\mathcal{C}}_{j}^{l}$.
  \begin{figure}[H]
\centering
\includegraphics[scale=0.5]{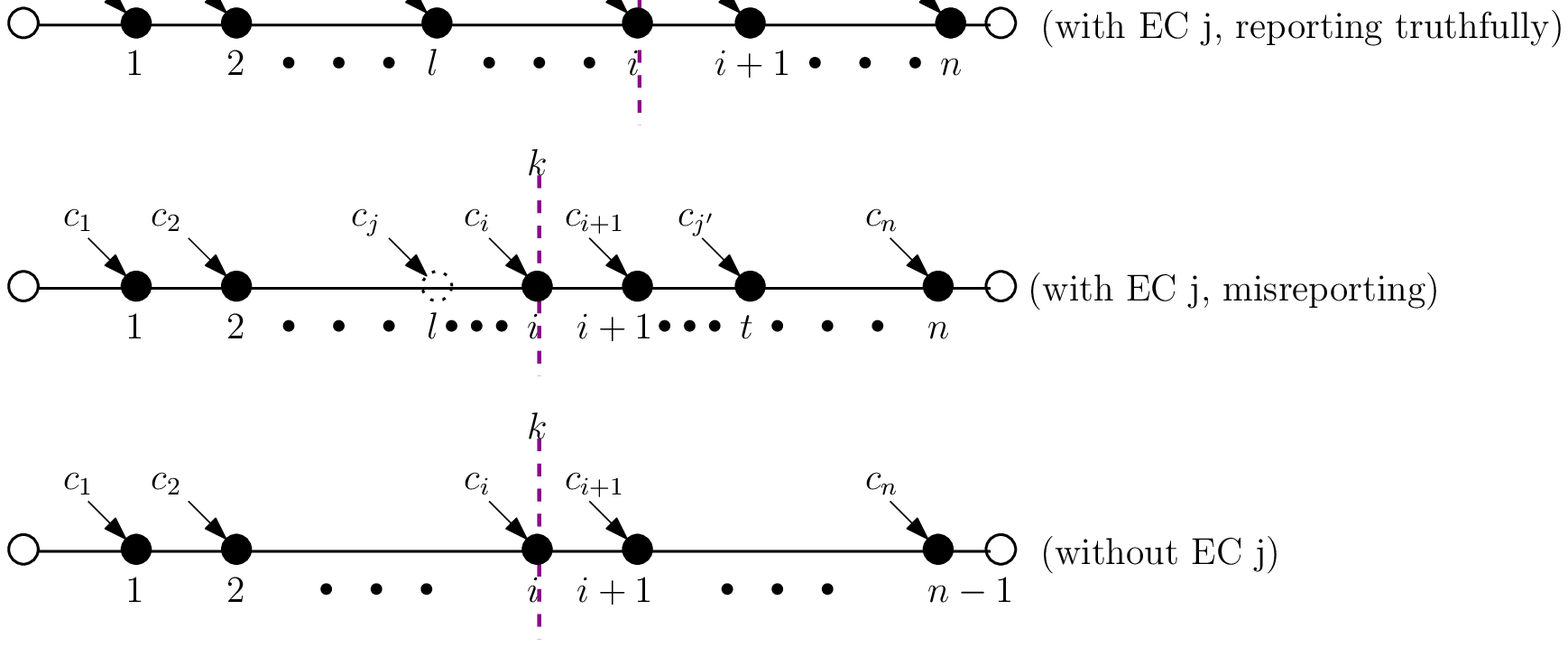}
\caption{Pictorial representation}
\label{theo8:8}
\end{figure}

Hence, considering \emph{Case 1} and \emph{Case 2} it can be concluded that EC $j$ does not gain by overbidding there true cost. In similar fashion, we can write the above mentioned equation for any position \emph{i} before \emph{k} and in the same way we can prove that $\boldsymbol{\mathcal{C}}_{j}^{i} \leq c_j$. Hence, $\max$ operator will still endure that if any agent $j$ deviates and wins, then his payment will be less than the true cost and hence deviation may not give any gain.
 



Hence, the theorem is proved.
 \end{proof}	

\section{Performance Evaluation}
We compared our proposed mechanisms against the benchmark mechanism ({\fontfamily{qcr}\selectfont random mechanism}). In this, the doctors are selected randomly and are paid their declared cost. We have utilized the \emph{coverage model} for the first fold of our hiring problem. The performance metric includes the {\fontfamily{qcr}\selectfont Interested doctors set size}, and {\fontfamily{qcr}\selectfont Number of doctors hired}. The unit of
cost and budget is \$.

\subsection{Simulation set-up}
For our simulation purpose, a social graph is generated randomly using {\fontfamily{qcr}\selectfont Networkx} package of python. It consists of {\fontfamily{qcr}\selectfont 1000 nodes (doctors)} and approximately {\fontfamily{qcr}\selectfont 28,250 edges}. The maximum and minimum degree a node can have is 
10\% and 1\% of the total available nodes respectively. The cost of each node as initial adapter is uniformly distributed over $[30,50]$, the cost of consultancy is uniformly distributed over $[35,50]$, and quality is uniformly distributed over $[20,50]$. The budget is considered in range $[100,1000]$.

\subsection{Result analysis}
\begin{figure}[H]
\begin{minipage}{.25\textwidth}
  \includegraphics[scale=.22]{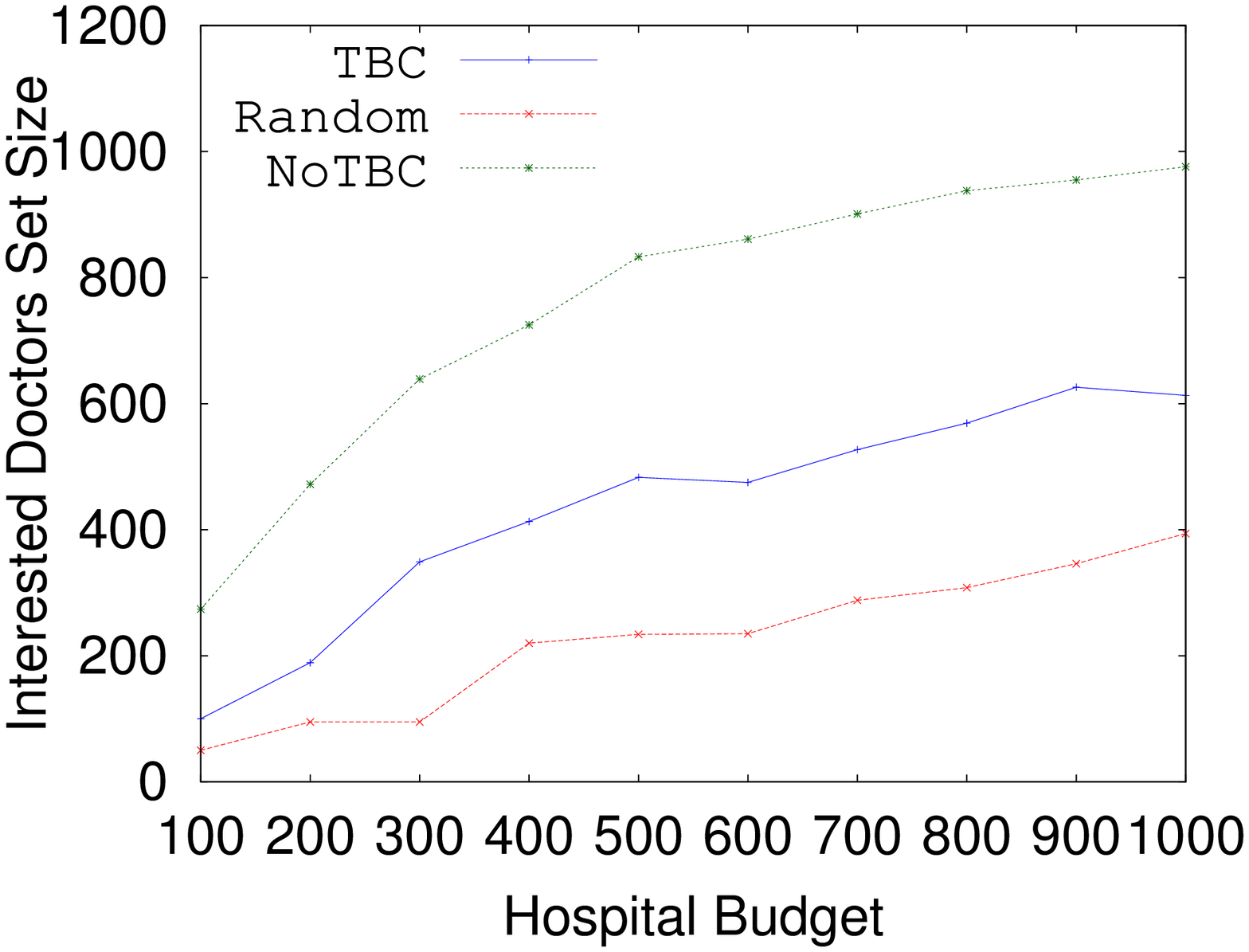}
  \caption{Interested doctors set size with budget $\boldsymbol{\mathcal{B}} \in [100,~1000]$}
  \label{fig:e1}
\end{minipage}%
\begin{minipage}{.25\textwidth}
  \includegraphics[scale=.22]{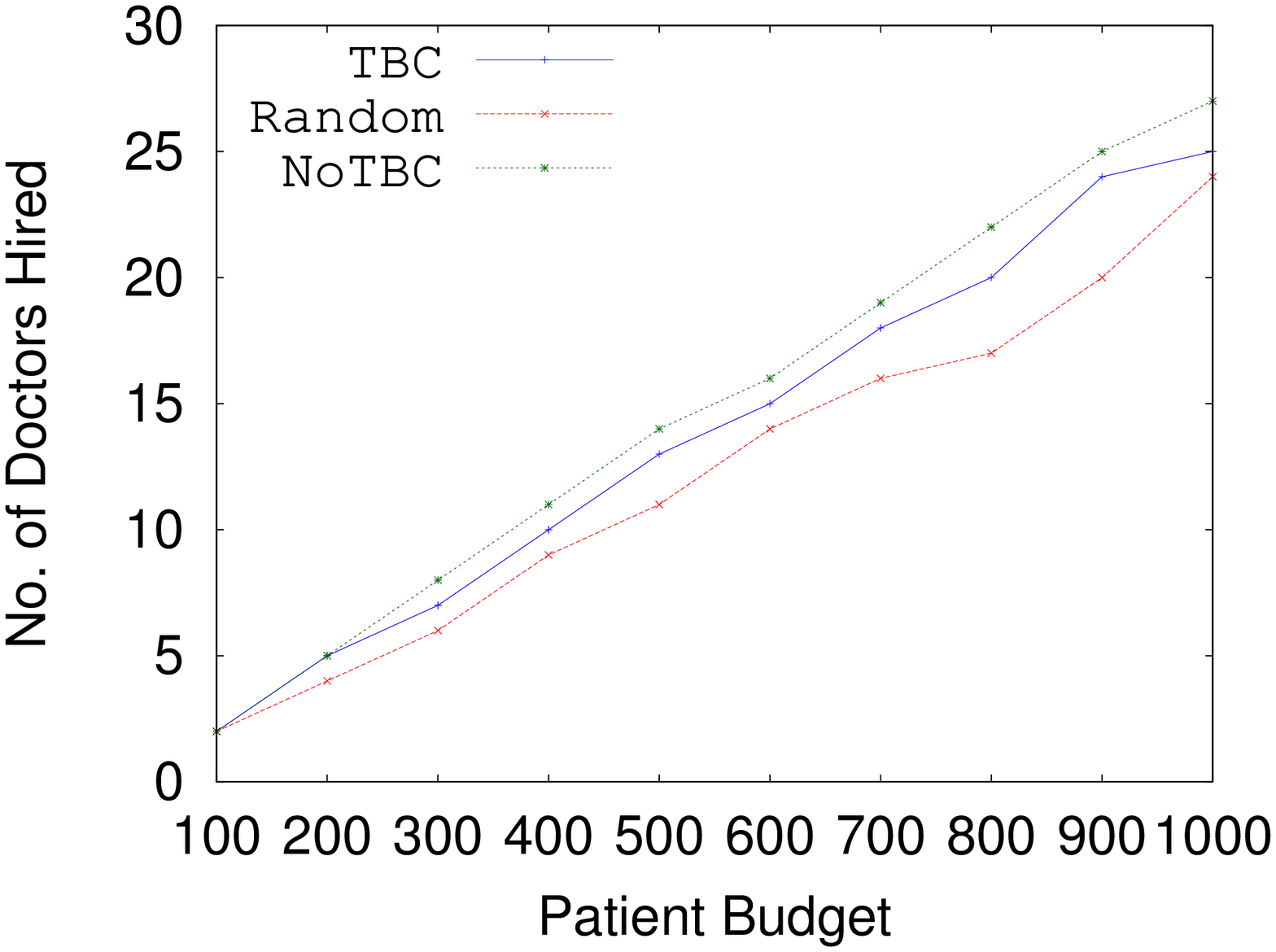}
  \caption{Hired doctors  with budget $\boldsymbol{\mathcal{B}'} \in [100,~1000]$}
  \label{fig:e2}
\end{minipage}
\end{figure}

 The simulation results shown in Figure \ref{fig:e1} shows the comparison of the interested doctors set size $i.e.$ the number of doctors acting as leaders and the number of doctors informed by the leaders about the hiring concept.     
\noindent It is seen in Figure \ref{fig:e1} that the interested doctors set size in case of NoTBC mechanism is higher than TBC mechanism and random mechanism. This nature of NoTBC mechanism is obvious due to the fact that the mechanisms (NoTBC-LI and NoTBC-DS) are utilizing almost the complete quota of the available budgets whereas TBC mechanism is utilizing only a part of total budget. With the increase in budget, one can easily see the increasing gap between NoTBC mechanism and TBC mechanism.
It can be seen evidently in Figure \ref{fig:e2} that the number of doctors hired in case of  NoTBC mechanism is higher than TBC mechanism and random mechanism. Similar reasoning can be given as above.

\section{Conclusions and future works}
In this paper, we have addressed the problem of hiring the doctors from outside of the hospital when a patient is constrained by budget.
Designing a more general mechanism in this environment for the set-up consisting of multiple patients say \emph{n} (each patient is associated to different hospitals) and \emph{m} doctors can be thought of as the future work.
\section*{Acknowledgment}
We would like to thank Prof. Y. Narahari and members of the Game Theory Lab. at IISc Bangalore; Ministry of Electronics and Information Technology (MeitY) Government of India and Media Lab Asia (YFR fellowship grant under Vivsvesvaraya PhD scheme), and Government of India Ministry of Human Resource Development (Institute scholarship).

\balance
\bibliographystyle{IEEEtran}
\bibliography{phd}

\begin{thebibliography}{10}
\providecommand{\url}[1]{#1}
\csname url@samestyle\endcsname
\providecommand{\newblock}{\relax}
\providecommand{\bibinfo}[2]{#2}
\providecommand{\BIBentrySTDinterwordspacing}{\spaceskip=0pt\relax}
\providecommand{\BIBentryALTinterwordstretchfactor}{4}
\providecommand{\BIBentryALTinterwordspacing}{\spaceskip=\fontdimen2\font plus
\BIBentryALTinterwordstretchfactor\fontdimen3\font minus
  \fontdimen4\font\relax}
\providecommand{\BIBforeignlanguage}[2]{{%
\expandafter\ifx\csname l@#1\endcsname\relax
\typeout{** WARNING: IEEEtran.bst: No hyphenation pattern has been}%
\typeout{** loaded for the language `#1'. Using the pattern for}%
\typeout{** the default language instead.}%
\else
\language=\csname l@#1\endcsname
\fi
#2}}
\providecommand{\BIBdecl}{\relax}
\BIBdecl

\bibitem{Guerriero_F_2011}
F.~Guerriero and R.~Guido, ``Operational research in the management of the
  operating theatre: A survey,'' \emph{Journal of health care management
  science}, vol.~14, no.~1, pp. 89--114, March 2011.

\bibitem{cardoen_b_2010}
B.~Cardoen, E.~Demeulemeester, and J.~Belien, ``Operating room planning and
  scheduling: A literature review,'' \emph{European journal of operational
  research}, vol. 201, no.~3, pp. 921--932, March 2010.

\bibitem{Berrada_1996}
I.~Berrada, J.~A. Ferland, and P.~Michelon, ``A multi-objective approach to
  nurse scheduling with both hard and soft constraints,'' \emph{Socio-economic
  planning sciences}, vol.~30, no.~3, pp. 183--193, 1996.

\bibitem{Huele_CV_2014}
C.~V.~Huele and M.~Vanhoucke, ``Analysis of the integration of the physician
  rostering problem and the surgery scheduling problem,'' \emph{Journal of
  medical systems}, vol.~38, no.~6, pp. 1--16, May 2014.

\bibitem{Starren_JB_2014}
J.~B. Starren, T.~S. Nesbitt, and M.~F. Chiang, ``Telehealth,'' \emph{Journal
  of biomedical informatics}, pp. 541--560, 2014.

\bibitem{Vikash2015EAS}
V.~K. Singh, S.~Mukhopadhyay, N.~Debnath, and A.~M. Chowdary, ``Auction aware
  selection of doctors in e-healthcare,'' in \emph{Proceeding of $17^{th}$
  Annual International Conference on E-health Networking, Application and
  services (HealthCom)}.\hskip 1em plus 0.5em minus 0.4em\relax Boston, USA:
  IEEE, 2015, pp. 363--368.

\bibitem{DBLP:journals/corr/SinghM16}
V.~K. Singh, S.~Mukhopadhyay, and R.~Das, ``Hiring doctors in e-healthcare with
  zero budget,'' in \emph{Advances on P2P, Parallel, Grid, Cloud and Internet
  Computing: Proceedings of the $12^{th}$ International Conference on P2P,
  Parallel, Grid, Cloud and Internet Computing (3PGCIC-2017)}, F.~Xhafa,
  S.~Caball{\'e}, and L.~Barolli, Eds.\hskip 1em plus 0.5em minus 0.4em\relax
  Cham: Springer International Publishing, 2018, pp. 379--390.

\bibitem{DBLP:journals/corr/SinghMSR17}
V.~K. Singh, S.~Mukhopadhyay, A.~Sharma, and A.~Roy, ``Hiring expert
  consultants in e-healthcare: {A} two sided matching approach,'' \emph{CoRR},
  vol. abs/1703.08698, 2017.

\bibitem{Chan:2016:PCP:2936924.2936967}
H.~Chan and J.~Chen, ``Provision-after-wait with common preferences,'' in
  \emph{Proceedings of the International Conference on Autonomous Agents and
  Multiagent Systems}, ser. AAMAS '16.\hskip 1em plus 0.5em minus 0.4em\relax
  Richland, SC: IFAAMAS, 2016, pp. 278--286.

\bibitem{Cardoen_B_1_2010}
B.~Cardoen, E.~Demeulemeester, and E.~Frank, ``Operating room planning and
  scheduling problems: A classification scheme,'' \emph{International journal
  of health management and information}, vol.~1, no.~1, pp. 71--83, 2010.

\bibitem{KO_Y_W_2017}
Y.~W. Ko, D.~H. Kim, S.~Uhmn, and J.~Kim, ``Nurse scheduling problem using
  backtracking,'' \emph{Advanced science letters}, vol.~23, no.~4, pp.
  3792--3795, April 2017.

\bibitem{Carter_2001}
M.~W. Carter and S.~D. Lapiere, ``Scheduling emergency room physicians,''
  \emph{Health care management science}, vol.~4, no.~4, pp. 347--360, December
  2001.

\bibitem{Dexter_F_2004}
F.~Dexter and A.~Macario, ``When to release allocated operating room time to
  increase operating room efficiency,'' \emph{Anesthesia and analgesia},
  vol.~98, no.~3, pp. 758--762, March 2004.

\bibitem{7920909}
R.~M. Pottayya, J.~C. Lapayre, and E.~Garcia, ``An adaptive videoconferencing
  framework for collaborative telemedicine,'' in \emph{Proceedings $31^{st}$
  International Conference on Advanced Information Networking and Applications
  (AINA)}, Taipei, Taiwan, March 2017, pp. 197--204.

\bibitem{TETC_2014_2386133}
C.~Tekin, O.~Atan, and M.~V.~D. Schaar, ``Discover the expert: Context-adaptive
  expert selection for medical diagnosis,'' \emph{IEEE transactions on emerging
  topics in computing}, vol.~3, no.~2, pp. 220--234, June 2015.

\bibitem{Singer:2012:WFI:2124295.2124381}
Y.~Singer, ``How to win friends and influence people, truthfully: Influence
  maximization mechanisms for social networks,'' in \emph{Proceedings of the
  $5^{th}$ International Conference on Web Search and Data Mining}, ser. WSDM
  '12.\hskip 1em plus 0.5em minus 0.4em\relax New York, NY, USA: ACM, 2012, pp.
  733--742.

\bibitem{Singer:2014:BFM:2692359.2692366}
------, ``Budget feasible mechanisms,'' in \emph{Proceedings of $51^{st}$
  Annual Symposium on Foundations of Computer Science (FOCS)}, Oct 2010, pp.
  765--774.

\bibitem{Khuller:1999:BMC:332530.332554}
S.~Khuller, A.~Moss, and J.~S. Naor, ``The budgeted maximum coverage problem,''
  \emph{Information processing letters}, vol.~70, no.~1, pp. 39--45, April
  1999.

\end{thebibliography}

\end{document}